\documentclass[
 twocolumn,
 amsmath,
 amssymb,
 aps, 
 physrev,
]{revtex4-2}

\usepackage{graphicx}
\usepackage{dcolumn}
\usepackage{bm}
\usepackage[pdftex,table]{xcolor}
\usepackage{hyperref}

\usepackage{enumitem}
\usepackage[colorinlistoftodos]{todonotes}

\hypersetup{
  colorlinks=true,
  citecolor=blue,
  linkcolor=blue,
  urlcolor=blue
}

\begin{document}

\preprint{APS/}

\title{\textbf{Feasibility of a collider-based detector search for upward-going fermions produced from gravitationally-bound dark matter within the Earth} 
}%

\author{Tamas Almos Vami}
\email{tamasvami@ucsb.edu}
\author{Sanjit Masanam}
\email{smasanam@ucsb.edu}
\affiliation{University of California Santa Barbara, Santa Barbara, CA, USA}

\author{Matt Bellis}
\email{mbellis@siena.edu}
\author{Josephine Swann}
\affiliation{Siena University, Loudonville, NY, USA}

\author{Philip Tanedo}
\author{Adam Green}
\affiliation{University of California Riverside, Riverside, CA}

\date{\today}

\begin{abstract}
Dark matter is theorized to be gravitationally captured within the Earth and to subsequently annihilate into dark photons. The dark photons kinetically mix with the Standard Model photons which would allow the decay of the dark photon to pairs of observable fermions near Earth's surface. We examine the feasibility and sensitivity of an LHC-class general-purpose detector to observe muons originating from these dark matter models. We calculate the expected signal rates at a hypothetical general-purpose detector and examine the effect of significant backgrounds on sensitivity. We also consider experimental efficiencies of such a detector to determine expected limits for the proposed analysis. We find these expected limits constitute a significant improvement over existing ones, strongly motivating a search for upward-going muons stemming from dark matter bound within the Earth.
\end{abstract}


\maketitle


\section{Introduction}
\label{sec:intro}

Accumulation of dark matter (DM) in celestial bodies have been a topic of interest for decades. The theoretical motivation is that a sufficiently high DM-nucleus interaction cross-section leads the DM to dissipate energy and become gravitationally bound to celestial objects~\cite{Press:1985ug,Silk:1985ax,Krauss:1985aaa,Griest:1986yu,Gaisser:1986ha}. The quantitative framework for this capture, and for the subsequent evolution of the captured population, was established in Refs.~\cite{Gould:1987ir,Gould:1987ju,Gould:1987tw,Gould:1991hx}. DM signatures from diverse astrophysical bodies have been probed, such as the Sun, white dwarfs, Jupiter, and other objects \cite{Gould:1999je, Leane:2022hkk, Phoroutan-Mehr:2025hjz, Boddy:2022knd,Li:2022wix,Leane:2021tjj,Graham:2018efk,Leane:2024bvh}.

With many prior studies focusing on DM signatures from far-away celestial bodies, the issue of hard-to-estimate astrophysical backgrounds becomes significant and decreases the sensitivity of these searches. The same interactions that allow DM to be captured by astrophysical objects also imply that it can be captured within the Earth, and a signature originating in the Earth avoids many of these astrophysical backgrounds. One may further theorize models in which DM interacts through a light mediator, such as a dark photon ($A'$)~\cite{Kobzarev:1966qya,Okun:1982xi,Holdom:1985ag,Holdom:1986eq,Fabbrichesi:2020wbt}. The dark photon interacts with the Standard Model (SM) only through its kinetic mixing with the SM photon, so it is very weakly coupled and can propagate through the Earth before decaying. With enough density, the trapped DM annihilates to dark photons, and the massive eigenstate of the photon--dark-photon mixture decays to a lepton pair near the surface of the Earth~\cite{Feng:2015hja}. Though upward-going fermions has been studied in some detail previously~\cite{habig1999neutrinoinducedupwardgoingmuonssuperkamiokande}, a pair of upward-going fermions is difficult to replicate via standard astrophysical backgrounds, motivating a search for pairs of fermions originating from within the Earth.

The realization that a light mediator opens up dilepton signatures of captured DM, rather than only the neutrino signatures considered in the 1980s work above, was made in Refs.~\cite{Schuster:2009au,Schuster:2009fc,Meade:2009mu}. Ref.~\cite{Feng:2015hja} extended this treatment to include the Sommerfeld enhancement of the annihilation rate, and explores the detection potential of the \texttt{IceCube} experiment~\cite{IceCube:2016zyt} to pairs of fermions and potential limits on weakly-interacting massive particle (WIMP) mass, dark photon mass, and the kinetic couplings. \texttt{IceCube} has instrumented roughly a cubic kilometer of ice and so its detection volume is very large. However, fermion pairs that arrive with very little angular separation are difficult to resolve in \texttt{IceCube}. Additionally, Ref.~\cite{Feng:2015hja} and other studies in the literature~\cite{Hamed:2009, Pospelov2008, Finkbeiner_2009} mainly consider the electron decay mode of DM annihilation. In this paper, we focus on the muons decay channel due to it being a lesser studied signature with appreciable rates that collider-based detectors are sensitive to. 

In this paper, we explore the feasibility of using detectors designed for collider experiments to detect and resolve pairs of upward-traveling muons arising from these dark photon decays. In Section~\ref{sec:theory}, we discuss the theoretical motivation for our phenomenological analysis and define the models we consider. In Section~\ref{sec:EarthShineGenerator}, we discuss the \textit{EarthShine} package which allows for the simulation of DM signal kinematics and detector acceptance. With this information, we detail our calculations of the expected signal rates for various parameters of the model in Section~\ref{sec:feasibilityDetector}. We also discuss the impact of significant backgrounds and of the general-purpose detector on the proposed analysis in Section~\ref{sec:feasibilityDetector}. Finally, we present expected limits of the search in Section~\ref{sec:expLims} and discuss our results and conclusions in Section~\ref{sec:Discussion}. 

\section{Theoretical Motivation}
\label{sec:theory}

We focus our study on two DM models. The first model, which we refer to as the \textit{core} model, has its capture and annihilation details presented in 2015 in Ref.~\cite{Feng:2015hja}. In this model, DM can interact with nuclei in the Earth, transferring energy to those nuclei and slowing down relative to the motion of the Earth. If DM loses enough energy that it can be gravitationally bound to the Earth, it will collect at near the core. Over the 4.5~billion-year lifetime of the Earth, DM population could reach its maximum (equilibrium) value and thus, the maximal signal rate for a given set of theoretical parameters.

The second model~\cite{Leane:2022hkk}, which we refer to as the \textit{floating} model, is similar to the \textit{core} model but theorizes that more of the DM population sits near the Earth's surface and thus is less densely concentrated at the core. Due to the lower density of DM, the annihilation rates of DM particles are possibly lower in this model. However, the more uniform distribution of DM in the Earth results in a wider range of decay lengths that produce decays near the detector. Due to these compensating effects, a significant number of events are expected at the detector in the \textit{floating} model case as well.

\subsection{Signal Event Rate Calculation}

Having presented an overview of the model, we now present our method of calculating the rate of muon pairs that arrive at the detector. Later, in Sections~\ref{sec:EarthShineGenerator} and~\ref{sec:feasibilityDetector}, we calculate the detector acceptance, efficiency, final yields, and also comment on the experimental resolution required to resolve both muons separately.

The rate at which the DM accumulates in the Earth, the rate of DM/anti-DM annihilations, and the lifetime of the dark photon are interrelated and determined by the following dark sector parameters: the mass of the DM candidate ($m_\chi$), the mass of the dark photon ($m_{A'}$), the kinetic mixing parameter between the standard model photon and the dark photon ($\epsilon$), and the dark fine structure constant ($\alpha_\chi$). Ref.~\cite{Feng:2015hja} investigates the range of parameters that would produce measurable rates in a detector with a volume of \texttt{IceCube}, one cubic kilometer. That work focuses on electron-positron pairs and produces upper-limit contours in $m_\chi$, $m_{A'}$, and $\epsilon$. Two values of $\alpha_\chi$ are explored there: one motivated by the DM density at thermal freeze-out ($\alpha_\chi^{\text{th}}$) and the other from constraints on distortions in the cosmic microwave background radiation ($\alpha_\chi^{\text{CMB}}$). Mass ranges are explored for the DM candidate from  $m_\chi = 100$~GeV $- 10$~TeV, for the dark photon $m_{A'} = 0.01 - 10$~GeV, and values for the kinetic mixing parameter $\epsilon = 10^{-11} - 10^{-6}$, and contours are identified in this parameter space where the rate inside of \texttt{IceCube} would be between 1-1000 events per 10 years, assuming 100\% detection efficiency.

The previous work focuses on dark photons that decay to electron-positron pairs inside of the detector volume. In this paper, we assume that the decays happen \textit{outside} the collider detector, in some larger volume of rock underneath the detector. However, only a subset of those will make it to the detector because of either energy loss or geometric acceptance.  Ref.~\cite{Green:2018qwo} provides a publicly available software package, \texttt{DarkCapPy}, for calculating expected capture and detection rates. By default, this tool calculates rates for \texttt{IceCube}. This rate calculation has since been merged into the larger \texttt{EarthShineGen} package~\cite{EarthShineGen}, which is what we use here and which provides both the expected yields and the signal kinematics within a single executable, as described in Section~\ref{sec:EarthShineGenerator} and Appendix~\ref{sec:codeavailability}. Using the generation volumes quoted in Section~\ref{sec:EarthShineGenerator}, we are able to scale by the generation \texttt{IceCube} volume ratio and acceptance fractions to estimate the rate for decays to muons at the general-purpose detector.

Throughout this section we set the dark fine structure constant to the CMB bound $\alpha_\chi^{\text{CMB}}=0.17(m_\chi/{\rm TeV})^{1.61}$ of Ref.~\cite{Feng:2015hja}. Similar to previous literature, we use this as a benchmark that fixes the scale of the rate, and we quote all results as functions of the model parameters rather than as absolute predictions. The benchmark grows with mass, and above $m_\chi\approx14$~TeV it exceeds $4\pi$, beyond which it is no longer a useful expansion parameter and the Born approximation underlying the capture and annihilation rates no longer applies. The sensitivity we quote is strongest well below this, at $m_\chi\approx3-4$~TeV. We retain this benchmark because it has a simple parametric form and because it allows a direct comparison with Ref.~\cite{Feng:2015hja}.

\begin{figure}[ht!]
    \centering
     \includegraphics[width=0.98\linewidth]{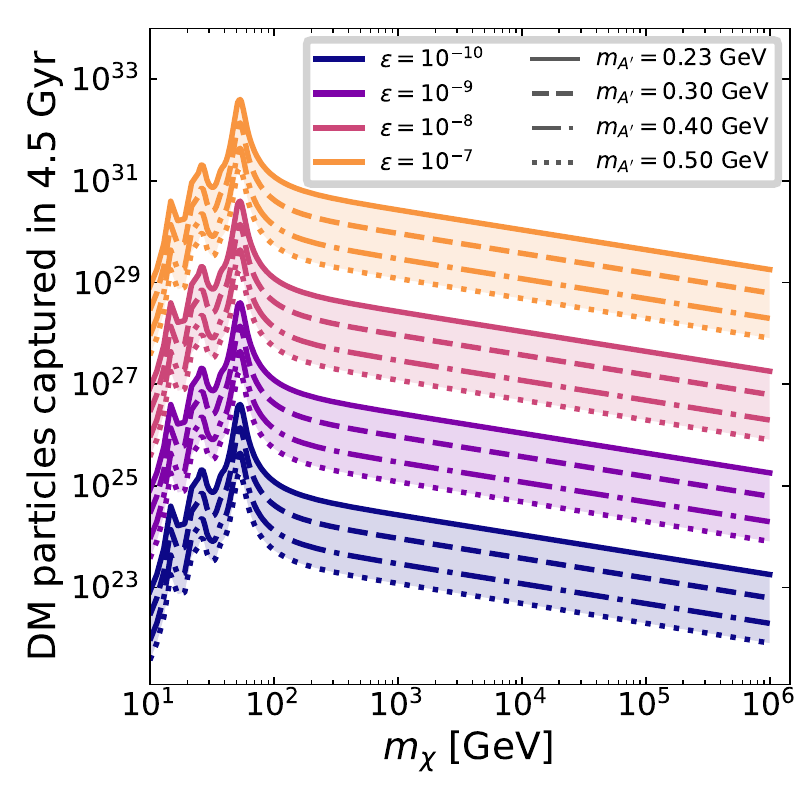}
    \caption{The number of DM particles that have collected in the Earth over 4.5 billion years, as a function of the DM mass. Color gives the kinetic mixing parameter $\epsilon$ and line style gives the dark photon mass $m_{A'}$, with the shaded band spanning the four dark photon masses at fixed $\epsilon$. The population scales as $\epsilon^{2}/m_{A'}^{4}$, so each color family is an exact rescaling of the others.}
    \label{fig:NumDMParticlesInEarth}
\end{figure}

The number of DM particles that have collected in the Earth over 4.5 billion years is shown in Figure~\ref{fig:NumDMParticlesInEarth}. Every curve peaks at approximately 50~GeV, which is where $m_\chi\approx m_{\text{Fe}}=52.5$~GeV and the DM scatters elastically with iron most efficiently. In this scatter, the DM particles essentially exchange their momenta with the at-rest (in Earth's frame) iron nuclei and become gravitationally trapped in the Earth. Since iron is the most abundant element within the core, a DM candidate with mass near $m_{\text{Fe}}$ would be more abundant in the Earth than for other masses. The same resonance for the lighter abundant elements produces the smaller peaks below 50~GeV, at the corresponding nuclear masses of silicon (26.3~GeV), magnesium (22.5~GeV) and oxygen (15.0~GeV).

\begin{figure}[ht!]
    \centering
    \includegraphics[width=0.98\linewidth]{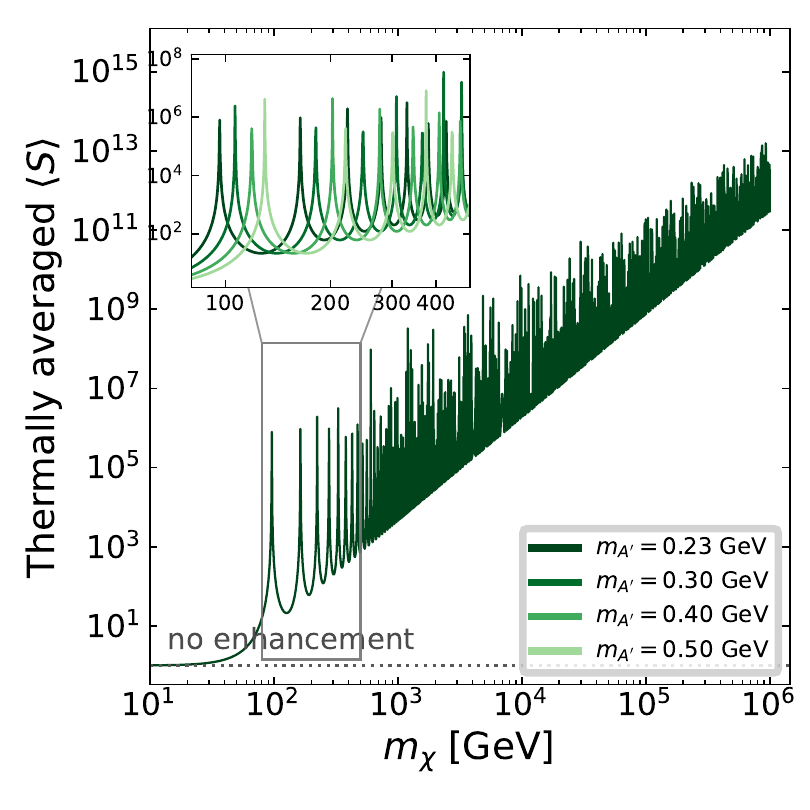}
    \caption{The thermally averaged Sommerfeld enhancement $\langle S\rangle$ as a function of the DM mass, with $\alpha_\chi=\alpha_\chi^{\text{CMB}}$. The quantity does not depend on $\epsilon$, so here color gives $m_{A'}$ rather than $\epsilon$. The spikes are the Sommerfeld resonances, which occur where $6\alpha_\chi m_\chi/(\pi^{2}m_{A'})$ passes through a perfect square. Because $\alpha_\chi^{\text{CMB}}$ itself grows with $m_\chi$, that combination runs as $m_\chi^{2.61}$ and the spacing between consecutive resonances narrows as $m_\chi^{-1.31}$. Above roughly 7~TeV they are closer together than the mass sampling of the figure, so what is drawn there is the envelope of the comb rather than individual resonances. For the same reason the four dark photon masses cannot be told apart over the full range, and the main panel shows only the reference $m_{A'}=0.23$~GeV. The inset zooms on 80 to 500~GeV, the widest window in which all four combs are still fully resolved, and where the first resonance is seen to move monotonically with $m_{A'}$, from 96~GeV at 0.23~GeV to 130~GeV at 0.50~GeV.}
    \label{fig:SommerfeldEnhancement}
\end{figure}

The thermally averaged annihilation cross section $\langle\sigma_{\rm ann}v\rangle$ before the Sommerfeld factor is applied needs no figure, since it is a single featureless power law over the parameter space considered here. It carries no dependence on $\epsilon$ at all, and depends on $m_{A'}$ only through $(m_{A'}/m_\chi)^{2}$, which is below one part in a thousand everywhere on this range, so all sixteen combinations of the four values of $\epsilon$ and the four dark photon masses coincide to within the width of a line. What remains is $\langle\sigma_{\rm ann}v\rangle\propto\alpha_\chi^{2}/m_\chi^{2}$, which with $\alpha_\chi=\alpha_\chi^{\text{CMB}}$ grows as $m_\chi^{1.22}$: it rises monotonically from about $4\times10^{-27}$ to $5\times10^{-21}$~cm$^{3}$/s between 10~GeV and 1~PeV, six orders of magnitude that are entirely the growth of $\alpha_\chi^{\text{CMB}}$ and not a feature of the annihilation itself. The Sommerfeld enhancement, by contrast, is structured, and is shown in Figure~\ref{fig:SommerfeldEnhancement}. It is likewise independent of $\epsilon$, and it multiplies $\langle\sigma_{\rm ann}v\rangle$ by many orders of magnitude, from of order $10^{2}$ at 100~GeV to of order $10^{9}$ at 100~TeV, with the narrow resonances reaching higher still.

\begin{figure}[ht!]
    \centering
    \includegraphics[width=0.98\linewidth]{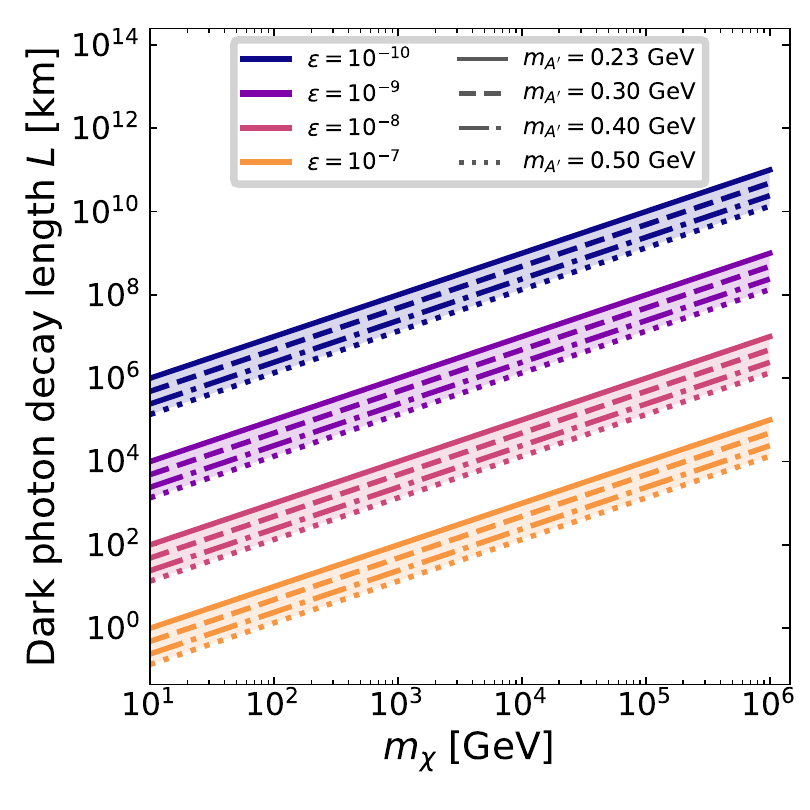}
    \caption{The lab-frame decay length $L$ of the dark photon as a function of the DM mass. Color gives $\epsilon$, line style gives $m_{A'}$ and the shaded band spans the four dark photon masses at fixed $\epsilon$, following the same convention as Figure~\ref{fig:NumDMParticlesInEarth}. The decay length scales as $1/\epsilon^{2}$, and within each color family as $B(A'\!\rightarrow\!\ell^{+}\ell^{-})/m_{A'}^{2}$, so the band spans a factor of 7.}
    \label{fig:decayLength}
\end{figure}

The decay length of the dark photon, which controls where in the Earth the visible decay occurs, is shown in Figure~\ref{fig:decayLength}. It scales as $1/\epsilon^{2}$, and has to be compared against two lengths that the figure does not show: a dark photon has to survive the passage out of the core and a distance of order $R_\oplus=6371$~km, and then decay inside the 4~km of rock below the detector from which we generate signal events. Both requirements are met only at the upper end of the $\epsilon$ range shown, since for $m_{A'}=0.23$~GeV the decay length falls from $10^{6}$~km at $\epsilon=10^{-10}$ to 1~km at $\epsilon=10^{-7}$ at $m_\chi=10$~GeV, and is larger by five orders of magnitude at $m_\chi=1$~PeV. Since one requirement is a lower bound on $L$ and the other an upper bound, only a narrow range of $\epsilon$ satisfies both at a given $m_\chi$, which is the origin of the two-sided exclusion in $\epsilon$ presented in Section~\ref{sec:expLims}.

\begin{figure}[ht!]
    \centering
    \includegraphics[width=0.98\linewidth]{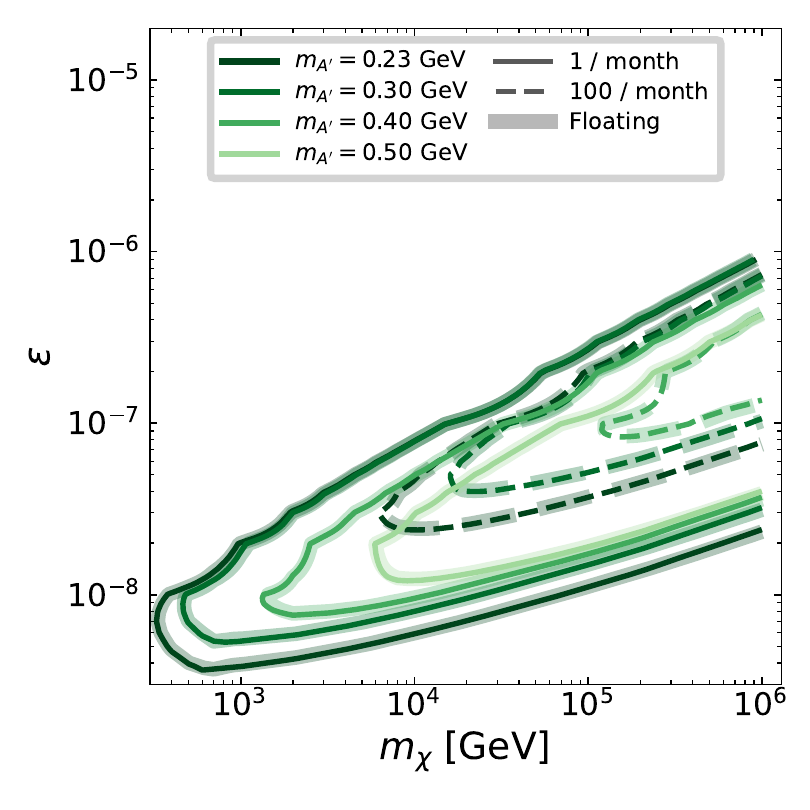}
    \caption{\textit{Core \& Floating model}. Expected number of upward-going muons per month in the $(m_\chi,\epsilon)$ plane, for muons arriving at the detector with an energy above 10~GeV, with $\alpha_\chi=\alpha_\chi^{\text{CMB}}$. Color gives the dark photon mass, on the same scale as Figure~\ref{fig:SommerfeldEnhancement}, and line style gives the rate level. The \textit{core} model is drawn as the line and the \textit{floating} model as the wide translucent band underneath it; the band tracks the line everywhere, which is the statement that the two models predict the same rate to within a few percent. The contours close on the left, at a few hundred GeV, but are cut off on the right by the edge of the scanned mass range at 1~PeV, and the 1 per month contour at $m_{A'}=0.23$~GeV is likewise cut off at the top by the edge of the scanned range in $\epsilon$ at $9\times10^{-7}$. The contours to the right of $m_\chi\approx14$~TeV lie where $\alpha_\chi^{\text{CMB}}$ exceeds $4\pi$, as discussed in Section~\ref{sec:theory}.}
    \label{fig:rateContours}
\end{figure}

For the \textit{core} model, we find that over much of the mass range considered in Ref.~\cite{Feng:2015hja}, the signal rate is far too low to be detectable and so we extend the DM mass range to span 1~TeV to 1000~TeV. This is a less-explored mass range, for three reasons. The expected number density is low and so is the event rate in direct-detection experiments. Above roughly 10~TeV the cosmological relic abundance can no longer be set by standard thermal freeze-out, so a more exotic production mechanism is required. And above 1~TeV there are now stringent astrophysical constraints on DM annihilation, from gamma-ray and cosmic-ray observations by H.E.S.S., HAWC, and LHAASO~\cite{Profumo:2017obk,Bell:2021pyy,Hiroshima:2025lqm}, which for secluded models of the kind considered here already exclude the simplest thermal relic cross section over much of this range. Those bounds constrain $\alpha_\chi$ rather than $\epsilon$, since they only require that the dark photon decay within the astrophysical source region, which is the case throughout our target region. They are therefore complementary to, and largely independent of, the search proposed here, and we return to this point in Section~\ref{sec:Discussion}. The contours are mapped out in Figure~\ref{fig:rateContours}. These rates are normalized to one month and we see that some parameter combinations would produce 10's or 100's of upward-traveling muons per month, reaching about 5500 per month at the best point, which lies at $m_\chi=1$~PeV and $\epsilon=3\times10^{-7}$ for $m_{A'}=0.23$~GeV. The rate contours highlight the strong dependence of the decay length on $\epsilon$. For large (small) $\epsilon$, the decay length is too large (small) to produce an appreciable number of decays near the detector. Thus, only a small range of $\epsilon$ produces appreciable rates at the detector. That range tracks upward in $\epsilon$ as $m_\chi$ increases, since a heavier DM particle gives a more boosted dark photon and hence a longer decay length at fixed $\epsilon$, which has to be compensated by a larger mixing. The accessible region also shrinks as $m_{A'}$ grows, closing off below $m_\chi\approx0.4$~TeV for $m_{A'}=0.23$~GeV and only below $m_\chi\approx6$~TeV for $m_{A'}=0.50$~GeV, and the peak rate falls from about 5500 to about 90 per month across that same range of $m_{A'}$. The \textit{floating} model rates are also shown in Figure~\ref{fig:rateContours}. They are barely distinguishable however, as they overlap almost entirely with the \textit{core} model rate contours: the floating generation volume is larger by a factor of $10^{4}$ while its geometric acceptance is smaller by the same factor, so the two predictions agree to about 4\%. The figure shows the contours for muons above 10~GeV at the detector; requiring instead that they arrive above 100~GeV gives contours that are indistinguishable from these by eye, because the muon spectrum is hard. Raising the threshold reduces the rate at the best point by only 23\% at $m_\chi=1$~TeV, by 6\% at 10~TeV and by 3\% at 100~TeV, which is a sub-percent shift of the contours. The exception is at the very bottom of the mass range, below the region shown, where $E_\mu\approx m_\chi/2$ falls under the cut itself: at $m_\chi=0.1$~TeV the rate above a 100~GeV threshold is zero, though this mass produces less than one muon per month at either threshold in any case.

The contours are drawn over the full scanned range, including the part above $m_\chi\approx14$~TeV where $\alpha_\chi^{\text{CMB}}$ leaves the perturbative regime discussed in Section~\ref{sec:theory}.

\section{EarthShine Generator}
\label{sec:EarthShineGenerator}

The \textit{EarthShine} generator is a  Monte Carlo simulation package developed to study the production and detection of upward-going fermions produced by DM annihilation within the Earth's interior. The software generates signal fermions according to a specified DM model and propagates the resulting particles through the surrounding rock and detector geometry. A discussion of energy loss parameterization and boost kinematics can be found in Appendix~\ref{app:eLossBoost}. The package allows for the estimation of signal kinematics and detector acceptance for a range of DM masses and mediator parameters. While the package allows simulation of the electron decay mode of DM annihilation, we do not consider it in this paper as we are assuming the decay happens in the rock below the detector. Electrons will lose energy far too quickly to reach the detector in appreciable quantities.

The generator is publicly released as \texttt{EarthShineGen}~\cite{EarthShineGen}, which merges the capture and annihilation rate calculation of \texttt{DarkCapPy}~\cite{Green:2018qwo} with the \textit{EarthShine} kinematics simulation, so that the absolute signal yields and the event kinematics are now obtained from one package rather than from two that have to be normalized to each other by hand. The detector geometry, the Earth model, and the model parameters ($m_\chi$, $m_{A'}$, $\epsilon$, $\alpha_\chi$) are all specified through a plain-text parameter card, and the events are written out in the Les Houches format. Further details are given in Appendix~\ref{sec:codeavailability}.

\subsection{Signal kinematics}

\begin{figure*}[!ht]
    \centering
    \includegraphics[width=0.32\linewidth]{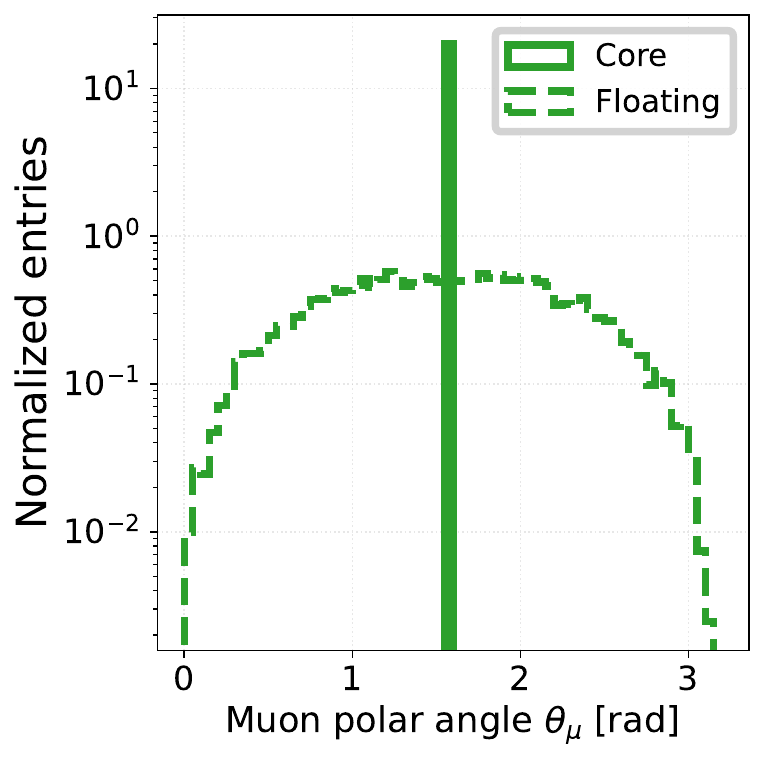}
    \includegraphics[width=0.32\linewidth]{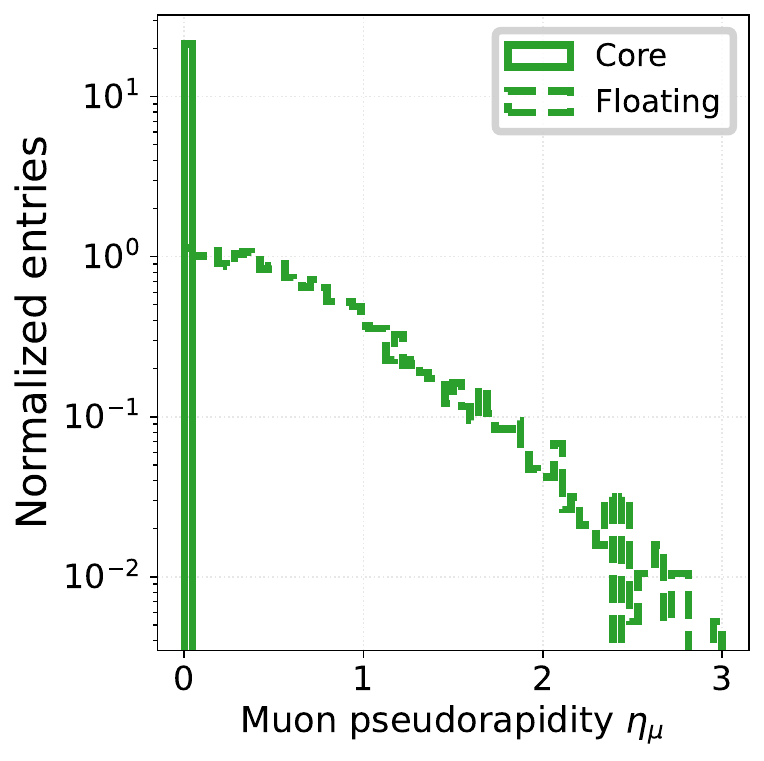}
    \includegraphics[width=0.32\linewidth]{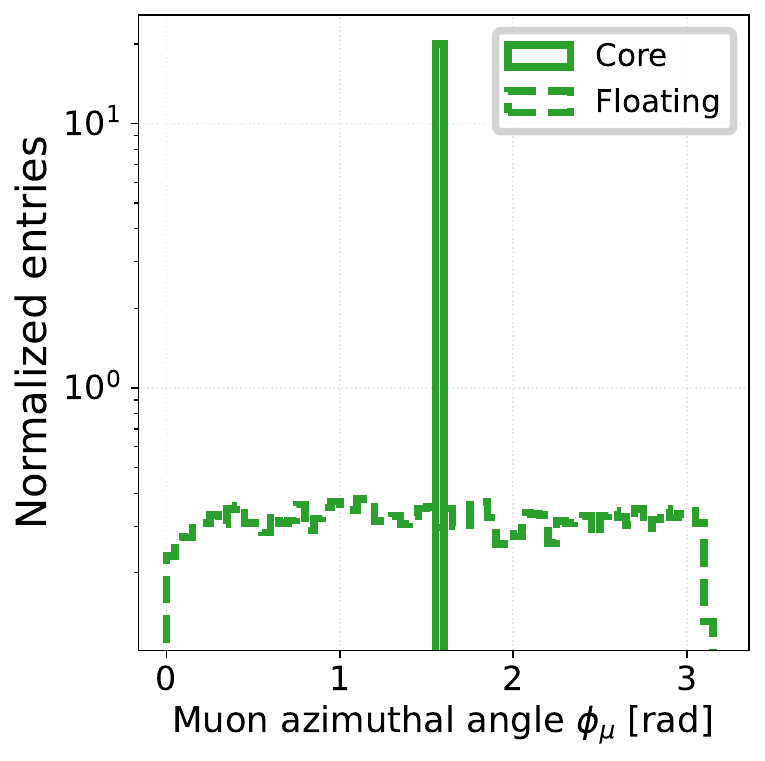}\\
    \includegraphics[width=0.32\linewidth]{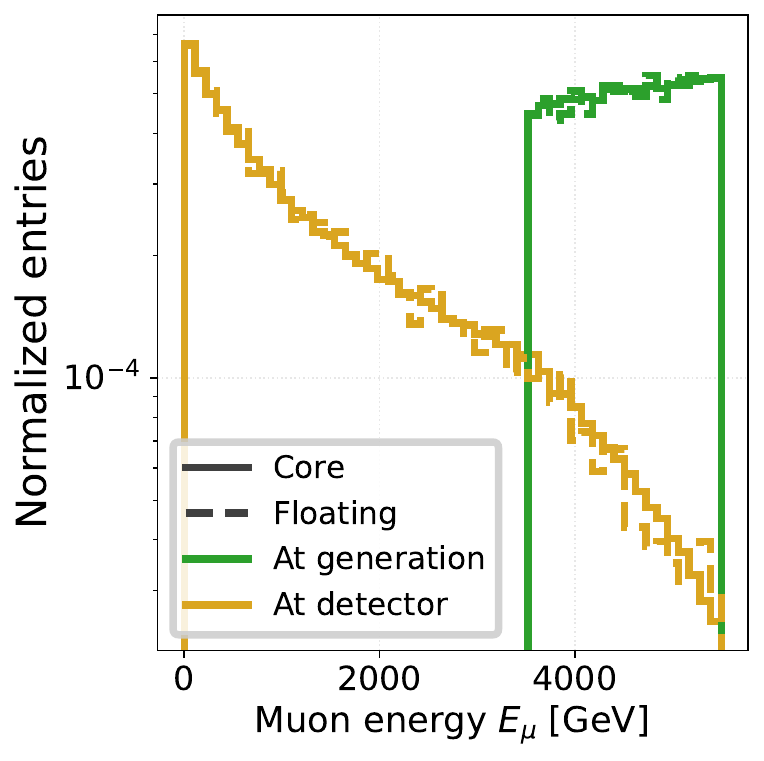}
    \includegraphics[width=0.32\linewidth]{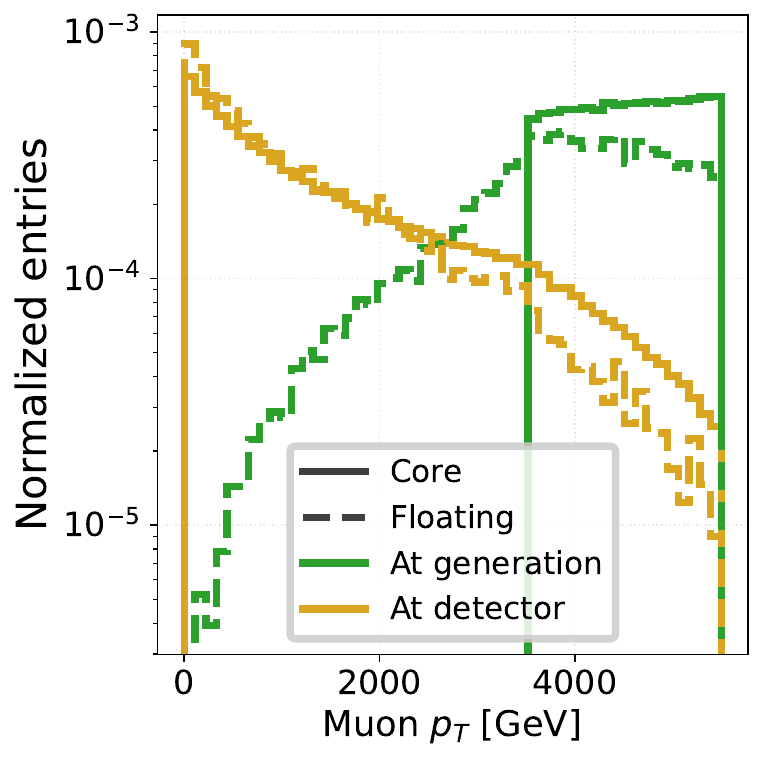}
    \caption{Energy, $p_T$, and angular distributions for muons coming from DM candidates with $m_{\chi}=10$~TeV. Two encodings are used simultaneously in every panel: line style gives the model, with the \textit{core} model solid and the \textit{floating} model dashed. In the lower row (energy and $p_T$), color additionally gives the stage at which the distribution is taken, with green the muons as generated at the decay point and gold the same muons at the detector, after energy loss in the rock is modeled; the angular panels of the upper row are taken at the detector and use a single color. In both models the muons are generated over a large volume of rock below the detector.}
    \label{fig:ePtAng}
\end{figure*}

The \textit{core} and \textit{floating} models are implemented separately in the \textit{EarthShine} package: each is a self-contained calculation, with its own DM density profile, its own generation volume, and hence its own signal kinematics and geometric acceptance. The user selects one of the two in the parameter card. In Figure~\ref{fig:ePtAng}, we show the simulated energy, $p_{T}$, and angular distributions at a detector near the surface for a $m_\chi=10~\text{TeV}$ candidate in both models. In general, we consider candidates within the mass range of 1~TeV to 1~PeV.

For the \textit{core} model, the DM particles are concentrated primarily in the Earth's core so the only muons that enter the detector are those produced going directly upward ($\theta = 180^\circ$). $\theta$ is defined as the angle from the vertical (z-axis). The energy of these muons can vary significantly, both at generation and at the detector. The boost the muons receive when they are produced causes their energy distribution to be smeared around a center of 5~TeV, half the DM mass. Both the energy loss of muons as they travel through rock and the initial smeared energy distribution at production cause the muon energy at the detector  to be much lower and far more spread out than at production. This same behavior is observed in the transverse momentum distributions since the upward traveling muons map entirely onto the transverse momentum plane as we have defined it. 

For the \textit{floating} model, the DM particle concentration is more uniform throughout the Earth and thus muons enter the detector at many angles. The energy distributions of muons both at generation and at the detector are similar to the \textit{core} model. This is expected as in both models, the muons pick up a significant boost from the dark photon and they experience energy loss as they travel through rock before entering the detector. The transverse momentum distributions are significantly different, a consequence of the muons entering at many different angles rather than directly upwards. The momentum of the muons no longer map entirely onto the transverse plane so the $p_{T}$ distribution has a lower mean than the energy distribution for this model. 

\begin{figure}[ht!]
    \centering
    \includegraphics[width=0.96\linewidth]{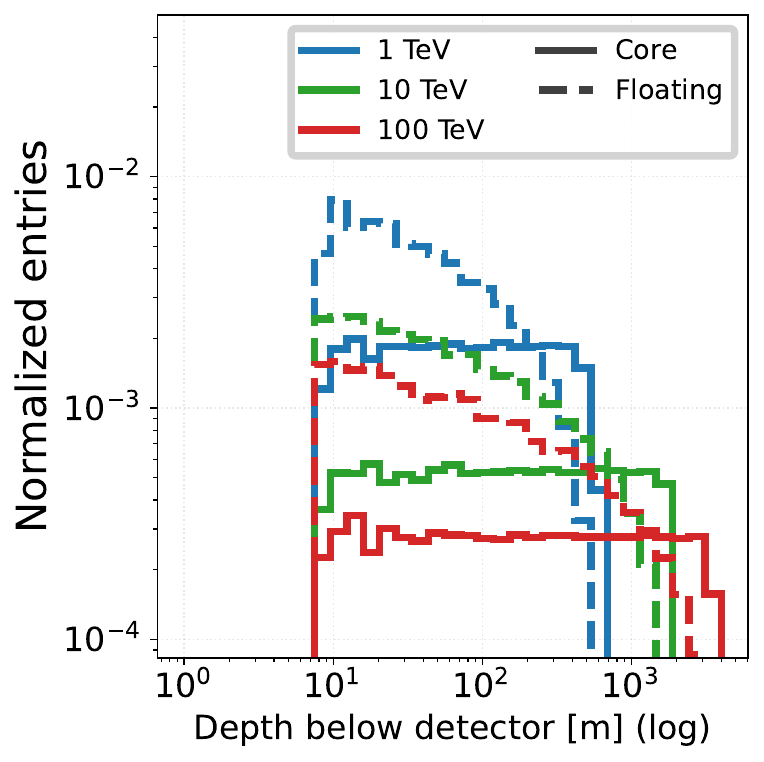}
    \caption{Distribution of the decay depth below the detector, shown on a logarithmic scale, for 1 TeV (blue), 10 TeV (green), and 100 TeV (red) DM candidates. The \textit{core} model is drawn with solid lines and the \textit{floating} model with dashed lines.}
    \label{fig:density_profiles}
\end{figure}

We assume the decays happen uniformly in the volume of rock below the detector. For heavier DM candidates, the signal muons generally carry more energy ($M(\chi) \approx E_{\mu}^1+E_\mu^2$) and thus can travel a longer distance through the rock (i.e. lose more energy) and still make it to the detector. Both models are shown in Figure~\ref{fig:density_profiles}, the \textit{core} model with solid lines and the \textit{floating} model with dashed lines. For the \textit{core} model, this relationship between energy and maximum decay depth is visible in the solid curves. However, for the \textit{floating} model, the events that reach the detector, for all 3 DM candidate masses, decay at a median depth of a few hundred meters, shallower than in the \textit{core} model. This is because muons can enter at $\theta\neq180^\circ$ in the \textit{floating} model. Thus, for a given decay depth, most muons travel a further distance ($d=z/\cos\theta$) than directly upward-going muons and so less events reach the detector due to energy loss.

\subsection{Detector Acceptance}
For our studies, we needed to choose the size of the detector. We considered choosing a geometry similar to IceCube due to its large detector volume. However, other detectors (CMS, ATLAS) possess stronger tracking capabilities and momentum resolution which allows one to isolate the signal signature not only with spatial information (upward vs. downward), but also using particle momentum. Thus, motivated by the geometries of CMS and ATLAS, we choose a cylinder that is 30 meters in length and 15 meters in diameter. 

\begin{figure}[ht!]
    \centering
    \includegraphics[width=0.98\linewidth]{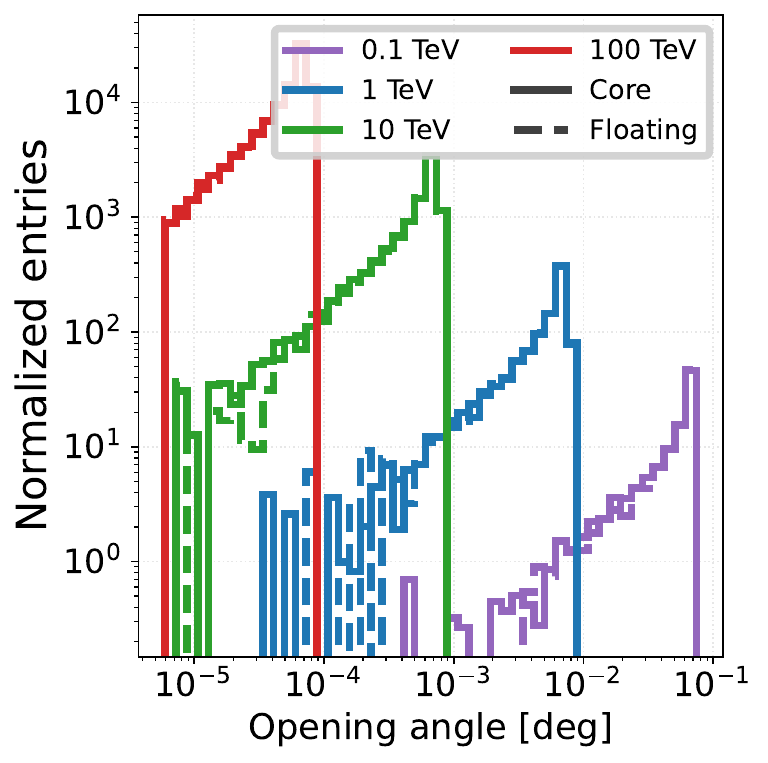}
    \caption{Opening angle distribution of muon pairs at the detector, for DM masses of 0.1, 1, 10, and 100~TeV (colors) and for the \textit{core} (solid) and \textit{floating} (dashed) models. At each mass the two models are nearly indistinguishable, because the opening angle is set by the internal kinematics of the dark-photon decay, which are identical between the models; it shifts by roughly a factor of ten for every factor of ten in mass, reflecting the $\theta_{\rm open}\sim 2\beta^{*}m_{A'}/E_{A'}$ scaling.}
    \label{fig:openAngleSep}
\end{figure}

Figure~\ref{fig:openAngleSep} shows, for several DM masses and for the \textit{core} and \textit{floating} models overlaid, the opening angle distribution of the muon pairs. At a fixed mass the two distributions are nearly identical between the models, and this is expected: the opening angle is governed by the internal kinematics of the dark-photon decay, which are the same in both cases. The opening angle between the two muons is set by the boost of the dark photon, $\theta_{\rm open}\sim 2\beta^{*}/\gamma_{A'} = 2\beta^{*}m_{A'}/E_{A'}$, where $\beta^{*}=\sqrt{1-r^{2}}$ is the muon velocity in the dark-photon rest frame and $r\equiv2m_\mu/m_{A'}$ measures how close the decay is to the dimuon threshold (Appendix~\ref{app:eLossBoost}). The factor $\beta^{*}$ is not negligible here: at $m_{A'}=0.23$~GeV the decay is only 9\% above the dimuon threshold, so $r=0.92$ and $\beta^{*}=0.39$, and the naive $2\,m_{A'}/E_{A'}$ would overstate the angle by a factor of $2.5$. This sets a maximum opening angle of $18~\mu$rad at 10~TeV, reached at $\theta_{\rm CM}=90^\circ$, with the distribution running down to zero for decays along the boost axis. Because the dark-photon mass $m_{A'}$ and its energy (fixed by the DM mass) are identical inputs in the two models, the opening-angle distribution does not depend on where the DM is concentrated or on the direction in which the muons travel. Because $E_{A'}$ scales with the DM mass, the opening angle scales as $\sim 1/m_\chi$, so the distribution shifts to smaller angles by roughly a factor of ten for every factor of ten in mass. For the 10~TeV candidate we find a mean opening angle of $6.0\times10^{-4}$~degrees with an RMS of $1.6\times10^{-4}$~degrees (about $10~\mu$rad) in both models. The transverse separation of the pair at the detector is this opening angle multiplied by the distance traveled from the decay point to the detector; because in both models the surviving events are dominated by decays occurring $\mathcal{O}(1~\mathrm{km})$ from the detector (Figure~\ref{fig:density_profiles}), with a median flight path of about $870$~m in both models, the flight-distance, and hence the separation, distributions coincide between the models as well, with a mean of $10.0$~mm and an RMS of $6.4$~mm at 10~TeV. The separation itself is shown in Figure~\ref{fig:msSeparation} of the following section, where it is compared against the much larger spread introduced by multiple scattering in the rock. The two models reach this same flight distance by different routes: in the \textit{core} model the muons travel vertically, so the flight path is the decay depth, while in the \textit{floating} model a median depth of about $390$~m is lengthened to a comparable path by the inclination of the tracks. The only quantities that differ between the two models, namely the direction of the incoming muons and the size of the generation volume, have no bearing on these two internal properties of the decay, which is why the distributions overlap. The separation of the pairs is very small (of order tens of mm or less), which means that although two muons are produced in a given candidate event, there is a high likelihood that they will be resolved as one upward going muon. In Section~\ref{sec:feasibilityDetector}, we will go more in depth on the spatial resolution required to resolve both upward going muons separately.

\subsubsection{Multiple scattering in the overburden}
\label{sec:multipleScattering}

The separations quoted above follow from the decay kinematics alone, with the muons propagated along straight lines. When propagating through a dense material, however, the muons do not travel in straight lines. They reach the detector through hundreds of meters of rock, the overburden, and multiple Coulomb scattering along that path is not a small correction to the opening angle of the decay. Taking the Highland parameterization~\cite{Highland1975}, in the form refit to Moli\`ere theory in Ref.~\cite{Lynch:1990sq} and quoted by the Particle Data Group~\cite{ParticleDataGroup:2024cfk}, integrated along the path so that the softening of the muon as it loses energy is accounted for,
\begin{equation}
    \sigma_\theta^2 = \left(1 + 0.038 \ln \frac{L}{X_0}\right)^{2} \int_0^L \left(\frac{13.6~\mathrm{MeV}}{\beta p(x)}\right)^{2} \frac{dx}{X_0},
    \label{eq:highland}
\end{equation}
with $X_0 = 26.54$~g\,cm$^{-2}$ for standard rock, a muon crossing the $\mathcal{O}(1~\mathrm{km})$ of overburden typical of the surviving events at 10~TeV picks up an RMS projected deflection of about $0.65$~mrad. The opening angle of the decay at the same mass is only $\sim 10~\mu$rad, so the scattering exceeds it by a factor of about 40, and the separation of the pair at the detector is set by the rock rather than by the dark photon.

\begin{figure}[ht!]
    \centering
    \includegraphics[width=0.98\linewidth]{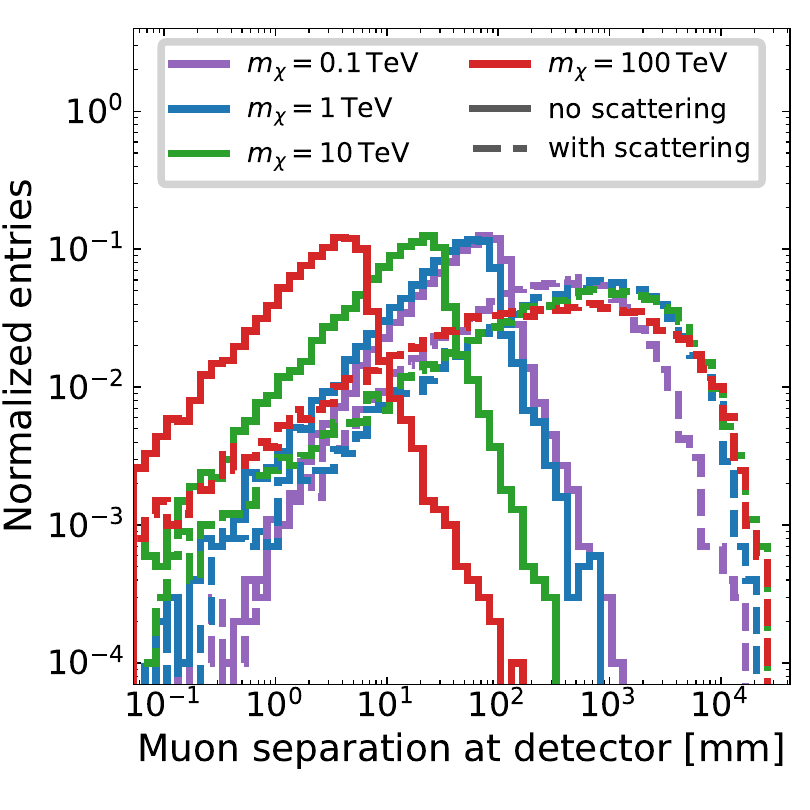}
    \caption{Separation of the two muons where they cross the detector, for the \textit{core} model and DM masses of 0.1, 1, 10, and 100~TeV (colors), without multiple scattering in the rock (solid) and with it (dashed). The \textit{core} and \textit{floating} models give separations that agree to roughly 10\%, both with and without scattering, so only the \textit{core} model is drawn; note that here, unlike in the other overlaid figures, the dashed lines are the scattered distributions and not the \textit{floating} model. Without scattering the separation follows the $1/m_\chi$ scaling of the opening angle; with scattering the mass dependence largely disappears and the median separation grows by more than an order of magnitude at every mass, from $1.4$~cm to $51$~cm at 10~TeV.}
    \label{fig:msSeparation}
\end{figure}

Figure~\ref{fig:msSeparation} shows the consequence of this scattering for the separation of the pair at the detector. Without scattering the separation falls with mass, following the $1/m_\chi$ scaling of the opening angle, and the 10~TeV distribution peaks in the tens of millimeters, consistent with the mean quoted above. The \textit{core} and \textit{floating} models give separations that agree to about 10\% at every mass, for the same reason their opening angles agree, so only the \textit{core} model is drawn. With scattering included the four mass points collapse onto nearly the same distribution, peaked between $0.2$ and $0.6$~m, because the deflection is governed by the path length and the momentum rather than by the parent kinematics. At 10~TeV the median separation moves from $1.4$~cm to $51$~cm, a factor of 36.

This reverses the conclusion drawn above. Where unscattered separations of tens of millimeters or less imply that a candidate event is very likely to be reconstructed as a single upward-going muon, the scattered distributions place the majority of pairs beyond half a meter, which the muon systems of the general-purpose LHC detectors resolve comfortably. The requirement placed on the detector in Section~\ref{sec:feasibilityDetector} is therefore considerably less demanding than the unscattered kinematics suggest, and the dimuon character of the signal survives the overburden.

Three caveats attach to this. Equation~\ref{eq:highland} describes the Gaussian core of the scattering distribution and not the single-hard-scatter tail, which at the $\sim10^{4}$ radiation lengths involved ($X_0 = 10.0$~cm in standard rock) is not a small correction, so the separations shown should be read as a lower bound on the spread. That same thickness also places us outside the range over which the parameterization was fit: Ref.~\cite{Lynch:1990sq} quotes an accuracy of 11\% for $10^{-3} \le x/X_0 \le 100$, and the overburden here is about two orders of magnitude beyond the upper end of that interval, so the logarithmic correction term is extrapolated rather than validated. We use it because no better closed form is available for this geometry, and because the conclusion drawn below survives a large error on the width. The scattering also degrades the reconstructed invariant mass of the pair, since the angle between the muons is no longer set by $m_{A'}$: at 10~TeV the median reconstructed mass moves from the generated $0.23$~GeV to above $1$~GeV. A search relying on a narrow dimuon resonance at the detector, rather than on the correlated upward-going pair, would have to account for this.

\begin{figure}[!ht]
    \centering
    \includegraphics[width=0.98\linewidth]{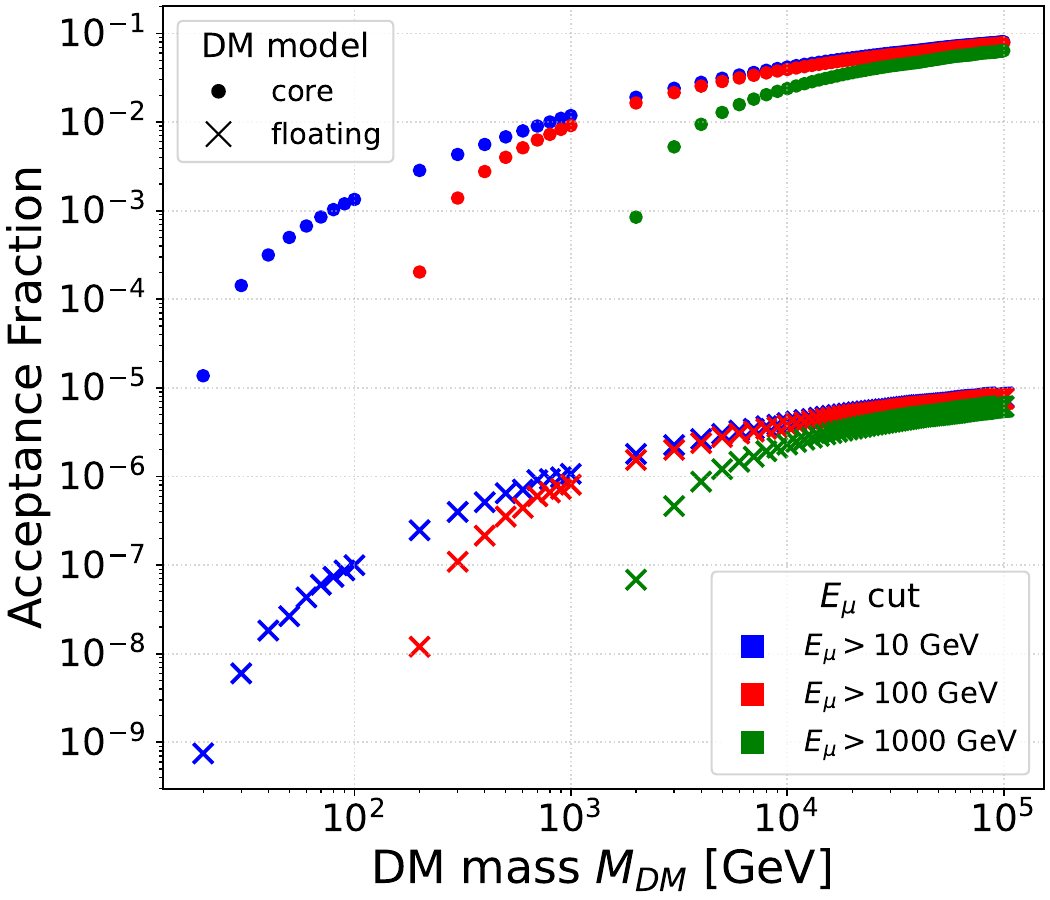}
    \caption{\textit{Core \& floating models.} Geometric acceptance for muons coming from DM candidates of different masses assuming the muons intersect the outer barrel of the detector. Various energy cuts are factored into the geometric acceptance (legend).}
    \label{fig:detectorAcceptanceCoreFloating}
\end{figure}

Figure~\ref{fig:detectorAcceptanceCoreFloating} shows the proportion of muons that intersected the detector as a function of the DM mass. For the \textit{floating} model, we generate the muons over a cylindrical volume below the detector: 4000 m in height (depth) and with a radius of 4000 m. This volume of rock is about 200 km$^3$ or about 200x larger than IceCube. For the \textit{core} model, the muons are traveling straight up and so we generate the muons over a cylindrical volume below the detector: 4000 m in height (depth) but with a radius of only 40 m. This volume of rock is about 0.02 km$^3$ or about 1/50$^{\rm th}$ the size of IceCube. 

These acceptance fractions have statistical uncertainty from the MC procedure of generating a finite number of muons below the detector and intersecting them with the detector. We calculate the binomial error on the acceptance fractions and determine that a 10\% nuisance on the expected signal rate is sufficient to capture this statistical uncertainty.

\section{Feasibility at a general-purpose LHC-class detector}
\label{sec:feasibilityDetector}

\subsection{Significant backgrounds}
\label{sec:sigBkg}

Having established a theoretical phase space where 10s to 100s of muons per month will reach and intersect the detector, we must consider any potential backgrounds which could pollute a search for the signal muons. Given that we are proposing this search to occur at a collider-based detector, we assume that the beams are off so that there are no particles originating from the collision point. In this case, the primary backgrounds are downward-traveling muons from cosmic rays, that might be misidentified as upward-traveling muons, and muons produced by neutrino interactions within the rock beneath the detector. For the purposes of this paper, we assume that the detector has the timing capabilities necessary to distinguish between upward and downward-traveling muons (more detailed discussion in Section~\ref{sec:timing}). Such a detector should allow for the rejection of all downward-going cosmic muons. However, timing measurement errors or fakes could result in the misidentification of a downward-going muon as upward-going. We assume that the general purpose detector can distinguish timing with 80\% efficiency. Additionally, the cosmic muon energy spectrum peaks at about 100~GeV and falls exponentially at energies larger than this~\cite{Achard_2004}. With this in mind, we anticipate low to zero contamination of the high-energy, upward-going muon signal region from the cosmic muon background.

The background muons from neutrino interactions requires consideration across a broad spectrum of energies, as though the signal muons produced by the decay of the dark photon initially have energies of 1~TeV and above, they will lose energy as they travel through the rock. However, the neutrino-induced muons are subject to the same energy loss as signal muons and the neutrino energy spectrum peaks in the GeV range~\cite{Vitagliano_2020}. Thus, we expect that a large-momentum cut will remove most if not all neutrino-induced background contamination.

The energy threshold is not the only handle on this background, and it is worth separating the two. A charged-current neutrino interaction in the rock produces a \textit{single} muon, whereas the signal is a correlated \textit{pair} of upward-going muons from the same dark photon decay. Two such muons arriving together requires either two coincident neutrino interactions, which is negligible at these rates, or a dimuon final state from charm production, which is a small branching fraction of an already small flux. Requiring two upward-going muons is therefore a rejection factor that does not depend on the muon energy, and Section~\ref{sec:multipleScattering} establishes that the pair is resolvable: multiple scattering in the overburden separates the two tracks by tens of centimeters at the detector, which the muon systems of general-purpose LHC detectors measure comfortably. This is what makes a threshold below 1~TeV worth considering at all, since the pair requirement survives where the momentum cut is relaxed. Quantifying it requires a dedicated simulation of the neutrino-induced rate, which we leave to future work.

\subsection{Detector Requirements}

The previous discussion has been independent of detector design (except for the outer geometry which played a role in the detector acceptance). However, to successfully carry out this search at a general-purpose LHC-class detector, a few requirements must be met. The two main backgrounds, as detailed in section~\ref{sec:sigBkg}, are rejected through a combination of sufficient timing and energy/momentum resolution. Additionally, to resolve the pair of signal muons separately requires sufficient spatial resolution. The resolutions required for these 3 aspects of a detector are discussed below.

\subsubsection{Spatial resolution}

As the solid curves of Figure~\ref{fig:msSeparation} show, the signal muon separation set by the decay kinematics alone varies significantly with respect to DM mass due to the boost picked up by the signal muons. On those separations alone the requirement would be severe: resolving both signal muons for mass points as high as 100~TeV would require a detector with a spatial resolution on the order of 1 $\mu$m, loosening to approximately 200 $\mu$m at the 100~GeV mass point, where 2 signal muons are correspondingly more likely to be separately resolved.

Multiple scattering in the overburden relaxes this requirement by orders of magnitude and largely removes its mass dependence, as discussed in Section~\ref{sec:multipleScattering}. Once scattering is included, the separations at the four mass points collapse onto nearly the same distribution, peaked between $0.2$ and $0.6$~m, so a two-track separation resolution of a few tens of centimeters resolves the majority of pairs at every mass considered (Figure~\ref{fig:msSeparation}). This is comfortably within the reach of the muon systems of general-purpose LHC detectors, so spatial resolution is not the limiting requirement for this search. At 100~TeV the signal muons are in any case already well separated in momentum from the potential backgrounds (Figure~\ref{fig:ptCoreFloating}), which offers a second handle for distinguishing signal muons from background muons of similar energy.

\subsubsection{Momentum resolution}

\begin{figure}[ht!]
    \centering
    \includegraphics[width=0.98\linewidth]{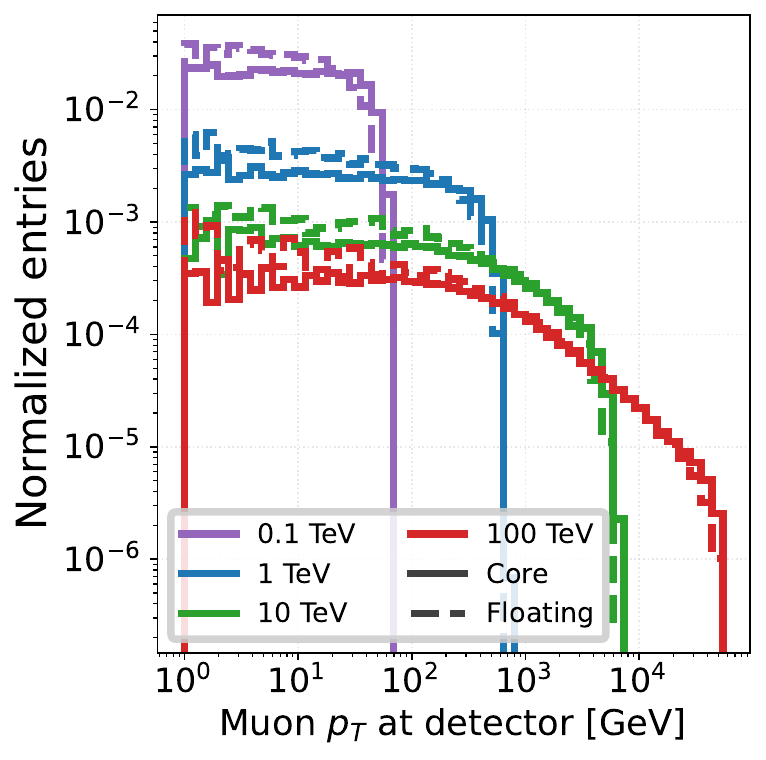}
    \caption{\textit{Core vs.\ floating models.} Muon transverse momentum at the detector (with energy loss) for DM masses of 0.1, 1, 10, and 100~TeV. Color encodes the DM mass, while the \textit{core} model is shown with solid lines and the \textit{floating} model with dashed lines.}
    \label{fig:ptCoreFloating}
\end{figure}

The transverse momentum of the signal muons that reach the detector sets the momentum resolution required to reconstruct them and to reject the softer background muons. Figure~\ref{fig:ptCoreFloating} compares the muon $p_T$ at the detector (after energy loss in the rock) for the \textit{core} and \textit{floating} models at DM masses of 0.1, 1, 10, and 100~TeV.

Due to energy loss, the $p_T$ distributions peak much lower than the initial DM mass point and fall until the initial mass point value. They also overlap significantly in the sub-TeV range. This indicates the detector must at minimum be able to resolve $p_T$ across the GeV to TeV range. Additionally, to determine the mass of a DM candidate, the detector must possess high, $\mathcal{O}(100~\mathrm{GeV})$, momentum resolution in the TeV range to accurately determine the $p_T$ of a high-energy signal muon.

\subsubsection{Timing resolution} \label{sec:timing}

\begin{figure}[!ht]
    \centering
    \includegraphics[width=0.98\linewidth]{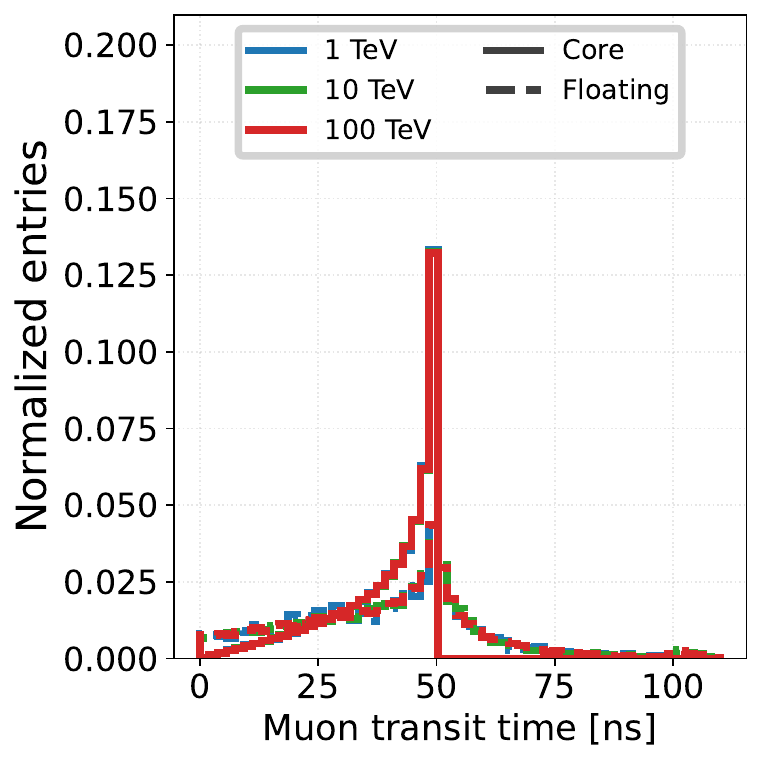}
    \caption{Time for a signal muon to cross the cylindrical detector, from its entry point to its exit point, for DM masses of $1$, $10$, and $100$~TeV (color). The \textit{core} model is drawn with solid lines and the \textit{floating} model with dashed lines. The curves for different masses nearly coincide because the muons are ultra-relativistic ($\beta\approx1$) at the detector, so the transit time is set by the crossing geometry rather than the DM mass. The \textit{core} distribution cuts off sharply at the detector diameter ($\approx50$~ns), while the inclined tracks of the \textit{floating} model extend the tail toward the full detector length ($\approx100$~ns).}
    \label{fig:transitTime}
\end{figure}

An important timing scale to consider is the time an individual signal muon spends inside the detector, i.e. the time to travel from its entry point to its exit point on the cylinder. The timing resolution must be comparable to this timescale to ensure upward-going signal muons are well separated from downward-going cosmic muons. Because the signal muons are ultra-relativistic at the detector, even at the $10$~GeV energy floor a muon has $\beta = \sqrt{1-(m_\mu/E)^2} \approx 0.99994$, and $1-\beta \lesssim 10^{-7}$ across the entire surviving population, this transit time is set almost entirely by the crossing geometry, $t \simeq \ell/c$ with $\ell$ the chord length through the cylinder, and is essentially independent of the DM mass. Figure~\ref{fig:transitTime} shows the transit-time distribution for DM masses of $1$, $10$, and $100$~TeV; the curves for the different masses lie on top of one another, confirming that the boost (and hence $\beta$) plays no role once the muon reaches the detector. The shape is instead governed by the incidence angle. In the \textit{core} model the muons travel vertically and cross the $15$~m diameter of the detector, producing a sharp edge at $\ell/c = 15~\mathrm{m}/c \approx 50$~ns. In the \textit{floating} model the muons arrive at a range of inclinations and acquire a substantial displacement along the cylinder (beam) axis, so the chord can approach the full $30$~m length of the detector; this smears the distribution out to a tail near $100$~ns while leaving the median transit time (${\sim}42$~ns) similar to the \textit{core} model.

\section{Expected limits}
\label{sec:expLims}

Here, we present expected limits of this proposed search for upward-going muons from DM within the Earth assuming zero and non-zero (1, 10, 100 events) high-energy muon contamination produced from neutrino interactions. We expect some background contamination but the zero background case is useful in understanding the maximum reach this proposed search could have. As discussed in section~\ref{sec:sigBkg}, we expect the downward-going cosmic muon background to be fully rejected while the neutrino-background will have $\mathcal{O}(1)$ event contamination over 1 year.

We use Combine~\cite{CMS:2024onh}, a public software package for statistical analysis, to calculate expected limits via a cut-and-count approach. As motivated in previous sections, we set a 10\% nuisance on the signal rate, an 80\% efficiency cut to model detector timing performance, and a 1 TeV energy cut to reject low energy backgrounds. Limits are set using a Cousins-Highland hybrid CLs approach~\cite{Cousins1992, Read_2002}.

\begin{figure}[!ht]
    \centering
    \includegraphics[width=0.98\linewidth]{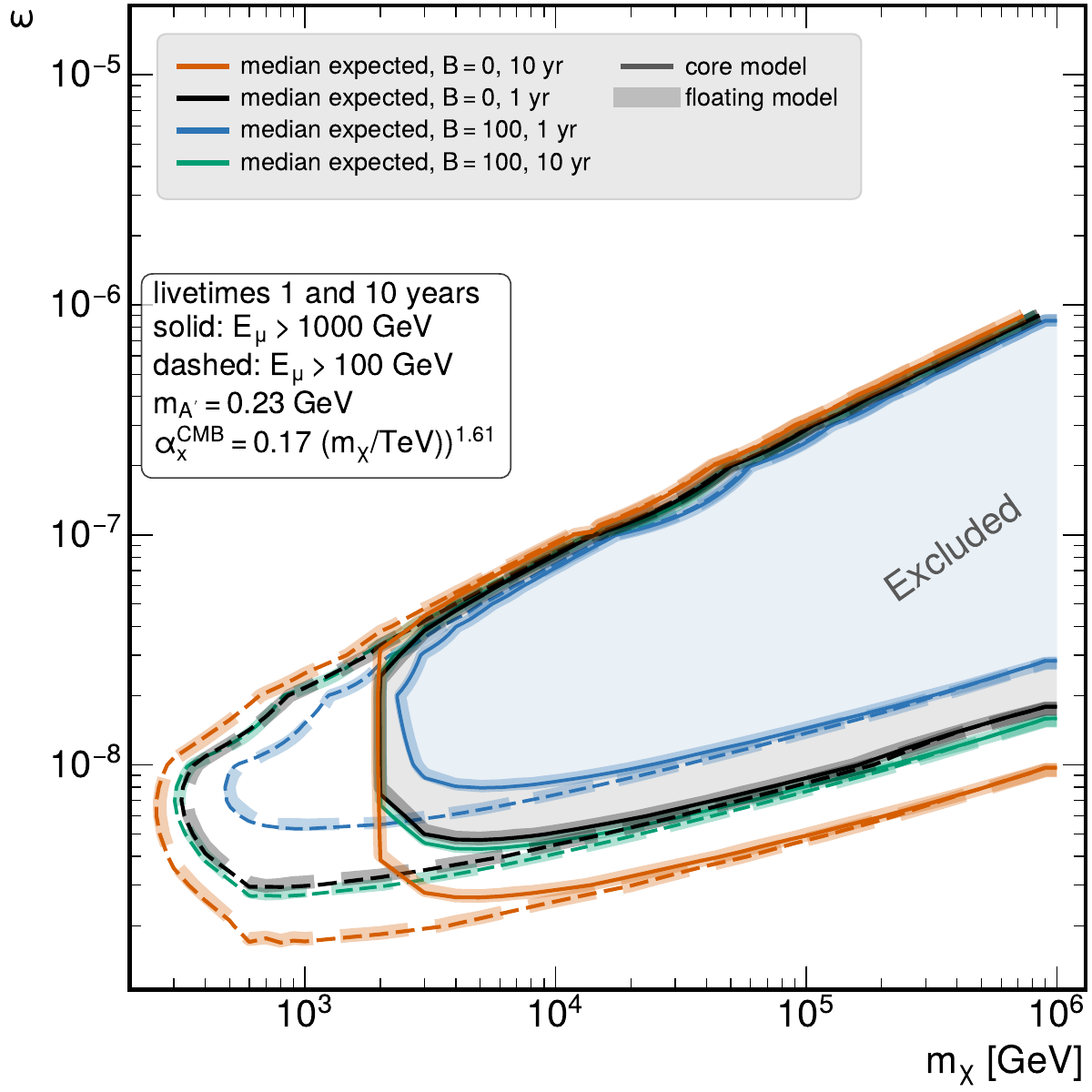}
    \caption{\textit{Core \& Floating model.} Expected exclusion in the $(m_\chi,\epsilon)$ plane, on the same axes as Figure~\ref{fig:rateContours}. Color gives the background hypothesis together with the live time, line style the muon energy threshold, and the \textit{core} model is the line with the \textit{floating} model the band beneath it. Only the median expected boundary is drawn. The shaded region is the exclusion for $B=0$ at 1 year with $E_\mu>1$~TeV; the other boundaries are drawn as lines on top of it.}
    \label{fig:coreFloatingExpLim}
\end{figure}

Figure~\ref{fig:coreFloatingExpLim} displays the expected reach of this search for $m_{A'}=0.23$~GeV and $\alpha_\chi^{\text{CMB}}=0.17(m_\chi/\mathrm{TeV})^{1.61}$, drawn on the same axes as the rate contours of Figure~\ref{fig:rateContours} so the two can be read against each other. The two-sided exclusion of $\epsilon$ for a given $m_\chi$ is a result of the decay length dependence on $\epsilon$, detailed in Section~\ref{sec:theory} and Figure~\ref{fig:decayLength}.

\begin{figure}[!ht]
    \centering
    \includegraphics[width=0.98\linewidth]{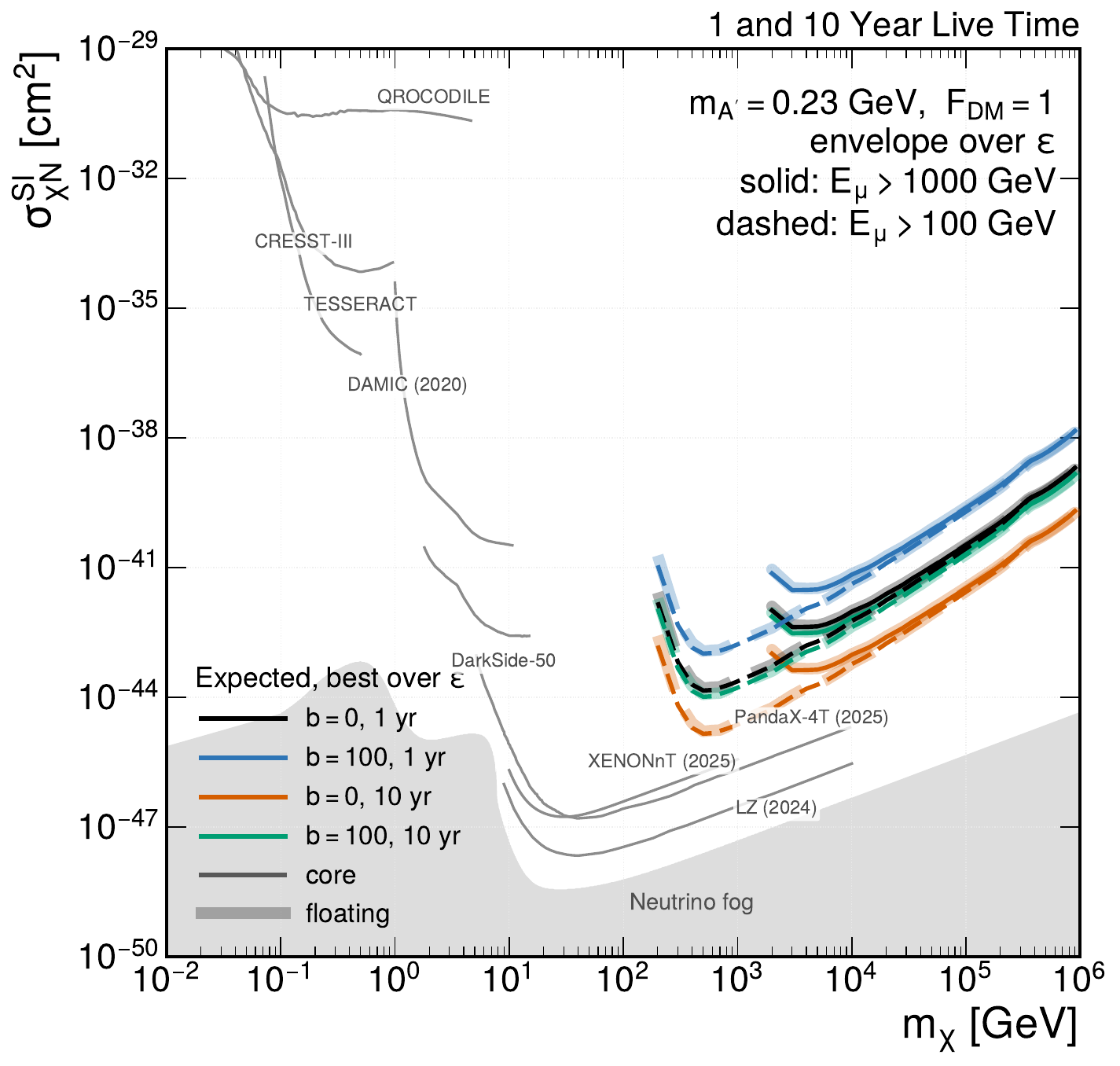}
    \caption{\textit{Core \& Floating model.} Expected limits on the spin-independent DM-proton cross section, against existing direct detection bounds~\cite{dd_limit_plot}. Color gives the background hypothesis and the live time, line style the muon energy threshold, and the \textit{core} model is the line with the \textit{floating} model the band beneath it. Curves are envelopes over $\epsilon$, and $B$ is the total yield in the exposure rather than a rate.}
    \label{fig:crossSectionLimits} 
\end{figure}

The figure varies three quantities at once. Color gives the background hypothesis together with the detector live time, for $B=0$ and $B=100$ events at 1 and at 10 years, following the same color code as Figure~\ref{fig:crossSectionLimits}. Line style gives the muon energy threshold, and the \textit{core} and \textit{floating} models are the line and the band beneath it. The limit cutoff at $m_\chi=2$~TeV in the solid curves is a consequence of the 1~TeV signal muon energy cut, since each muon carries about half the DM mass; the dashed curves, at a 100~GeV threshold, extend the excluded region down to $m_\chi\approx0.3$~TeV and open a second lobe at low mass that the higher threshold cannot reach.

Two features of the figure are worth reading carefully. The exclusion at $B=100$ after 10 years is wider than at $B=0$ after 1 year, so over this parameter space exposure buys more than perfect background rejection does. And $B$ is a total yield in the whole exposure rather than a rate, so the 10 year curves assume the same integrated contamination as the 1 year ones.

Figure~\ref{fig:coreFloatingExpLim} also recasts the ($m_\chi,\ \epsilon$) limits onto the spin-independent DM-proton cross section for the $B = 0$ and $B = 100$ cases. This is calculated using
\begin{equation}\label{eq:sigma_recast}
    \sigma_{\chi p} = \frac{16 \pi \epsilon^{2} \alpha \alpha_{\chi} \mu_{\chi p}^{2}}{m_{A'}^{4}},
\end{equation}
where $\alpha$ is the fine-structure constant, $\alpha_{\chi}$ the dark coupling, and $\mu_{\chi p}$ the
DM-proton reduced mass. Equation~\ref{eq:sigma_recast} is the heavy-mediator limit of Eq.~17 of
Ref.~\cite{Feng:2015hja}, and it applies here by a wide margin: the mediator mass squared is
$m_{A'}^{2} = 0.053~\text{GeV}^{2}$ while the momentum transfer in a nuclear recoil is
$2 m_{N} E_{R} \sim 10^{-7}~\text{GeV}^{2}$, so the propagator is flat over the whole relevant range of
momentum transfer and the dark matter form factor is unity.

For each DM mass, the excluded $\epsilon$ is obtained from Figure~\ref{fig:coreFloatingExpLim},
converted through Eq.~\ref{eq:sigma_recast}, and then minimized over $\epsilon$. Because $\epsilon$ fixes both the scattering cross section and the dark photon decay length, the signal rate is not a function of $\sigma_{\chi p}$ alone. We therefore quote the strongest cross section limit reached by the excluded $\epsilon$.

The comparison shown in Figure~\ref{fig:crossSectionLimits}, against existing direct detection limits sourced from~\cite{dd_limit_plot}, highlights the vast mass range this search covers. The limits run out to $10^{6}$~GeV, whereas the deepest direct-detection
limits stop at $10^{4}$ GeV (LZ) and $9.6 \times 10^{3}$ GeV (PandaX-4T); over most of the range
covered here there is no direct-detection curve to compare against at all.

The lower edge of the mass range is set by the energy threshold rather than by the rate, since each muon carries about half the DM mass. With the 1~TeV threshold used for the limits above, no signal survives below $m_\chi\approx2$~TeV; lowering the threshold to 100~GeV extends the reach down to $m_\chi\approx200$~GeV and moves the point of strongest sensitivity from 3~TeV to 0.5~TeV. The gain is therefore largely one of mass reach: at fixed mass the looser threshold improves the cross section by a factor of 4 at 3~TeV, falling to 1.6 at 10~TeV and 1.2 at 100~TeV, because the muon spectrum is hard and most of the signal already lies above 1~TeV. We quote the 1~TeV threshold as the headline because it is the one for which the background argument of Section~\ref{sec:sigBkg} is made quantitatively. The 100~GeV curves are shown because the dominant rejection of the neutrino-induced background there is the requirement of two correlated upward-going muons rather than the momentum cut, as discussed in that section, and that handle does not weaken as the threshold is lowered.

In terms of the cross section reached, the strongest median expected exclusion at the 1~TeV threshold and one year of live time is $4.2 \times 10^{-43}\,\mathrm{cm}^{2}$ for the \textit{core} model and $4.6 \times 10^{-43}\,\mathrm{cm}^{2}$ for the \textit{floating} model, both at $m_\chi\approx3-4$~TeV. This is comparable to the best reach of DarkSide-50
($2.5 \times 10^{-43}\,\mathrm{cm}^{2}$), obtained at a dark matter mass some two orders of magnitude
larger. Because the limit on the signal yield is fixed by the background hypothesis alone, the cross section reached scales inversely with the exposure: ten years of live time moves each curve down by a decade, to $4.2 \times 10^{-44}\,\mathrm{cm}^{2}$ at $B=0$. The LHC experiments have been recording cosmic ray data since 2008, so an exposure of this order is realistic. Notably, ten years with $B=100$ reaches a lower cross section than one year with no background at all, so for this search exposure is worth more than perfect background rejection.

In the narrow window where this search does overlap the dedicated xenon experiments, between
about 2 and 10 TeV, it is weaker by roughly 3 orders of magnitude. This is expected as sensitivity here comes from counting upward-going muons in a general-purpose collider detector, not from a dedicated low-background nuclear-recoil measurement.

Direct detection is not the only constraint on this parameter space, and Figure~\ref{fig:crossSectionLimits} should not be read as though it were. Two others apply over the mass range shown. Taking $\alpha_\chi=\alpha_\chi^{\text{CMB}}$, perturbativity restricts the model itself to $m_\chi\lesssim14$~TeV, as discussed in Section~\ref{sec:theory}. And the indirect-detection limits on DM annihilation above 1~TeV from H.E.S.S., HAWC, and LHAASO~\cite{Profumo:2017obk,Bell:2021pyy,Hiroshima:2025lqm} constrain $\alpha_\chi$ directly, and within this specific model are more competitive than the reach quoted here over much of the range. The comparison to direct detection is made because it is the most familiar way to gauge the strength of a DM search, not because it is the only bound in play.

\section{Discussion}
\label{sec:Discussion}

In this paper, we have considered the feasibility of conducting a search for multiple DM-candidates that are gravitationally bound within the Earth at collider-based detectors. We introduced a new simulation package, \textit{EarthShine}, to study the production and detection of upward-going muons produced from DM annihilation within the Earth. Using this package, we calculated the signal muon kinematics for various DM models and modeled detector acceptance of these signal muons. After considering the main background for this search and the detector requirements necessary for a successful search, we calculated the expected limits for the search and compared them to existing direct detection limits.

The expected limits of this phenomenological analysis show that regions of phase space previously unexplored by direct detection experiments will be excluded for only 1 year of live-time and even with significant (100 event) background contamination. Taking the dark photon model literally, the region that is both newly probed and internally self-consistent is the tip near $(m_\chi,\epsilon)\approx(10~\mathrm{TeV},3\times10^{-8})$, since above $m_\chi\approx14$~TeV the benchmark $\alpha_\chi^{\text{CMB}}$ leaves the perturbative regime. We nevertheless present the limits over the full scanned range because they are Born-approximation expressions that can be recast: for a model in which the corresponding coupling stays perturbative at these masses, the heavier region is a legitimate bound once our expressions are converted into that model's variables. Both CMS and ATLAS meet the detector requirements laid out in Section~\ref{sec:feasibilityDetector} and have live-times significantly longer than 1 year, having recorded cosmic ray data since 2008. Because the limit on the signal yield is set by the background hypothesis alone, ten years of live time lowers the reach by a decade, and ten years with 100 background events still reaches further than one year with none, so exposure is worth more to this search than perfect background rejection. Lowering the muon energy threshold from 1~TeV to 100~GeV would extend the mass reach down to $m_\chi\approx200$~GeV and move the point of strongest sensitivity to $m_\chi\approx0.5$~TeV, at the cost of the neutrino-induced background rejection that the higher threshold provides. While these detectors will likely have detector-related acceptance efficiencies and complications, a more sophisticated treatment of the background rejection could compensate any lost sensitivity from this idealized phenomenological study.

A distinguishing feature of the search proposed here is that it costs no additional hardware. It reuses the existing instrumentation of a general-purpose collider detector, with its fine spatial granularity, its magnetic momentum measurement, and its timing, during periods when the beams are off. This is a very different instrument from the sparse photomultiplier array of \texttt{IceCube}, whose sensitivity to this signal was estimated in Ref.~\cite{Feng:2015hja} under the assumption of 100\% detection efficiency and with a far cruder treatment of backgrounds than the one presented here. \texttt{IceCube} compensates with a detection volume roughly $50$ times larger than the \textit{core} model generation volume used here, but it cannot resolve the two muons of the pair, nor measure their momenta, nor exploit the hardness of the signal spectrum against the soft neutrino-induced background. The comparison is therefore one of volume against resolution, and the present study indicates that the resolution of a collider detector is worth a great deal.

Overall, the proposed search for upward-going muons from DM gravitationally bound within the Earth finds that a detector similar to CMS/ATLAS could probe and exclude significant phase space with a live-time as low as 1 year and background contamination as high as 100 events.

\appendix
\section{Code and Data Availability}
\label{sec:codeavailability}

The generator used for all signal samples in this work, \texttt{EarthShineGen}, is publicly available at \url{https://github.com/EarthShineGen/EarthShineGen}~\cite{EarthShineGen}. It is a single executable that merges the capture, annihilation, and dark photon decay rate calculation of \texttt{DarkCapPy}~\cite{Green:2018qwo} with the kinematics simulation, so that the expected yields and the signal kinematics are computed together. One change to the original \texttt{DarkCapPy} has been to add functionality to calculate the rates assuming the dark photon couples to muon pairs, rather than just electron pairs. For the branching fraction for $A' \to \mu^+\mu^-$ we use the results from Buschmann et al.~\cite{Buschmann:2015awa} The full chain, from the capture rate through the fraction of dark photons decaying inside the generation volume to the fraction of muon pairs that reach the detector, is reported by the generator itself, with the last factor measured from the thrown events rather than supplied externally. The dark matter model parameters, the Earth model, and the detector geometry are all set in a plain-text parameter card, so the package is not tied to a particular experiment, and events are written as Les Houches event files that can be passed directly to a detector simulation. The package ships a validation of its rate calculation against \texttt{DarkCapPy}, with which it agrees to one part in $10^{7}$ while running about three orders of magnitude faster.

The simulated event samples, the tabulated capture and annihilation rates, and the plotting scripts that produce every figure in this paper are archived at Ref.~\cite{PhenoPaperZenodo}. The record includes a figure index mapping each figure to the script and the input files it is drawn from, so that any figure here can be regenerated from the released data alone.

\section{Energy Loss \& Boost Kinematics in \textit{EarthShine}}\label{app:eLossBoost}

The energy loss of muons passing through rock within the Earth is explicitly parameterized with the following approximation to the Bethe-Bloch formula \cite{ParticleDataGroup:2024cfk},
\begin{equation}
    -\frac{dE}{dx}=a+bE,
    \label{eq:Eloss}
\end{equation}
which has the analytical solution,
\begin{equation}
    E(x)= \left(E_0 + \frac{a}{b}\right) e^{-bx} - \frac{a}{b}.
\end{equation}
Here, $a$ is the ionization loss parameter (with units of GeV/(g\,cm$^{-2}$)) and $b$ is the radiative loss parameter (with units of cm$^{2}$/g). These two symbols are used for the energy loss parameters in this appendix only; the decay kinematics below use $r$ and $\gamma_{A'}$. This parameterization is applied to the $\mathcal{O}(1) - \mathcal{O}(100)$~TeV energy regime.

Additionally, the DM annihilation results in the candidate mass being converted to the energy of the dark photon and because $m_\chi \gg m_{A'}$ in the phase space we are considering, the dark photon is highly boosted,
$$E_{A'} = m_\chi, \;\; \gamma_{A'} = \frac{E_{A'}}{m_{A'}} = \frac{m_\chi}{m_{A'}}, \;\; \beta_{A'} = \sqrt{1-\gamma_{A'}^{-2}}.$$
For instance, at $m_\chi=1$~TeV and $m_{A'}=0.23$~GeV, $\gamma_{A'}=4.3\cdot10^3$. The pair of signal muons decay back-to-back and isotropically in the rest frame of the dark photon,
$$E_\mu^{\rm CM} = \frac{m_{A'}}{2}, \quad p_\mu^{\rm CM} = \frac{m_{A'}}{2}\sqrt{1-r^2}, \quad r \equiv \frac{2m_\mu}{m_{A'}},$$
where $r$ measures how close the decay is to the dimuon threshold and $\beta^{*}=\sqrt{1-r^{2}}$ is the muon velocity in that frame, as used in Section~\ref{sec:EarthShineGenerator}. After boosting along the $A'$ direction,
$$E_\mu^\pm = \frac{m_\chi}{2}\left(1 \pm \beta_{A'}\sqrt{1-r^2}\cos\theta_{\rm CM}\right);$$

given the isotropic decays, the boost results in an approximately flat distribution for $E_\mu^\pm$ centered at $m_\chi/2$.

\section*{Acknowledgments}

We would like to thank  Joe Incandela, Liam Brennan, and Danyi Zhang for useful discussions. We acknowledge the use of Anthropic's Claude Sonnet and Opus models for code development and manuscript editing. All outputs were checked by the authors who take full responsibility for the content. We used the UCSB computational facilities administered by the Center for Scientific Computing at the California NanoSystems Institute and Materials Research Laboratory (an NSF MRSEC; DMR-1720256) and purchased through NSF CNS-1725797. UCSB is supported by the US Department of Energy under grant DE-SC0011702. Support for UCSB is also made possible by the Joe and Pat Yzurdiaga endowed chair in experimental science. PT is supported by a NSF CAREER award \#2045333. The Siena group is supported by a grant from the US National Science Foundation under Award No. EPP-2310056.

\newpage
\bibliography{apssamp}

@article{Buschmann:2015awa,
    author = "Buschmann, Malte and Kopp, Joachim and Liu, Jia and Machado, Pedro A. N.",
    title = "{Lepton Jets from Radiating Dark Matter}",
    eprint = "1505.07459",
    archivePrefix = "arXiv",
    primaryClass = "hep-ph",
    reportNumber = "MITP-15-036, IFT-UAM-CSIC-15-047, FTUAM-15-13",
    doi = "10.1007/JHEP07(2015)045",
    journal = "JHEP",
    volume = "07",
    pages = "045",
    year = "2015"
}

@misc{PhenoPaperZenodo,
    author       = "Vami, Tamas Almos and Masanam, Sanjit and Bellis, Matt",
    title        = "{Data and plotting scripts for: Feasibility of a collider-based detector search for upward-going fermions produced from gravitationally-bound dark matter within the Earth}",
    year         = "2026",
    publisher    = "Zenodo",
    doi          = "10.5281/zenodo.22164855",
    url          = "https://doi.org/10.5281/zenodo.22164855"
}

@misc{dd_limit_plot,
    author = "Mor\r{a}, K. D. and Olcina, I. and Chen, X. and Ma, Y.",
    title = "{dd\_limit\_plot: a compilation of direct-detection dark matter limits}",
    howpublished = "\url{https://github.com/kdund/dd_limit_plot}",
    year = "2026",
    note = "Accessed 2026-08-13"
}

@article{
    CMS:2024onh,
    author = "Hayrapetyan, Aram and others",
    collaboration = "CMS",
    title = "The {CMS} statistical analysis and combination tool: {\textsc{Combine}}",
    eprint = "2404.06614",
    archivePrefix = "arXiv",
    primaryClass = "physics.data-an",
    reportNumber = "CMS-CAT-23-001, CERN-EP-2024-078",
    year = "2024",
    journal = "Comput. Softw. Big Sci.",
    doi = "10.1007/s41781-024-00121-4",
    volume = "8",
    pages = "19"
}

@PREAMBLE{
 "\providecommand{\noopsort}[1]{}" 
 # "\providecommand{\singleletter}[1]{#1}%" 
}

@article{Li:2022wix,
    author = "Li, Lingfeng and Fan, JiJi",
    title = "{Jupiter missions as probes of dark matter}",
    eprint = "2207.13709",
    archivePrefix = "arXiv",
    primaryClass = "hep-ph",
    doi = "10.1007/JHEP10(2022)186",
    journal = "JHEP",
    volume = "10",
    pages = "186",
    year = "2022"
}

@article{Leane:2022hkk,
    author = "Leane, Rebecca K. and Smirnov, Juri",
    title = "{Floating dark matter in celestial bodies}",
    eprint = "2209.09834",
    archivePrefix = "arXiv",
    primaryClass = "hep-ph",
    reportNumber = "SLAC-PUB-17655, LTH-1320",
    doi = "10.1088/1475-7516/2023/10/057",
    journal = "JCAP",
    volume = "10",
    pages = "057",
    year = "2023"
}

@article{Vitagliano_2020,
   title={Grand unified neutrino spectrum at Earth: Sources and spectral components},
   volume={92},
   ISSN={1539-0756},
   url={http://dx.doi.org/10.1103/RevModPhys.92.045006},
   DOI={10.1103/revmodphys.92.045006},
   number={4},
   journal={Reviews of Modern Physics},
   publisher={American Physical Society (APS)},
   author={Vitagliano, Edoardo and Tamborra, Irene and Raffelt, Georg},
   year={2020},
   month=Dec }

@misc{habig1999neutrinoinducedupwardgoingmuonssuperkamiokande,
      title={Neutrino-induced upward-going muons in Super-Kamiokande}, 
      author={A. Habig},
      year={1999},
      eprint={hep-ex/9903047},
      archivePrefix={arXiv},
      primaryClass={hep-ex},
      url={https://arxiv.org/abs/hep-ex/9903047}, 
}

@article{Graham:2018efk,
    author = "Graham, Peter W. and Janish, Ryan and Narayan, Vijay and Rajendran, Surjeet and Riggins, Paul",
    title = "{White Dwarfs as Dark Matter Detectors}",
    eprint = "1805.07381",
    archivePrefix = "arXiv",
    primaryClass = "hep-ph",
    doi = "10.1103/PhysRevD.98.115027",
    journal = "Phys. Rev. D",
    volume = "98",
    number = "11",
    pages = "115027",
    year = "2018"
}

@article{Phoroutan-Mehr:2025hjz,
    author = "Phoroutan-Mehr, Mehrdad and Fetherolf, Tara",
    title = "{Probing Superheavy Dark Matter with Exoplanets}",
    eprint = "2503.00125",
    archivePrefix = "arXiv",
    primaryClass = "hep-ph",
    journal = "arXiv e-prints",
    month = "2",
    year = "2025"
}

@article{Boddy:2022knd,
    author = "Boddy, Kimberly K. and others",
    title = "{Snowmass2021 theory frontier white paper: Astrophysical and cosmological probes of dark matter}",
    eprint = "2203.06380",
    archivePrefix = "arXiv",
    primaryClass = "hep-ph",
    reportNumber = "FERMILAB-CONF-22-151-T",
    doi = "10.1016/j.jheap.2022.06.005",
    journal = "JHEAp",
    volume = "35",
    pages = "112--138",
    year = "2022"
}

@article{Leane:2021tjj,
    author = "Leane, Rebecca K. and Linden, Tim",
    title = "{First Analysis of Jupiter in Gamma Rays and a New Search for Dark Matter}",
    eprint = "2104.02068",
    archivePrefix = "arXiv",
    primaryClass = "astro-ph.HE",
    reportNumber = "SLAC-PUB-17594",
    doi = "10.1103/PhysRevLett.131.071001",
    journal = "Phys. Rev. Lett.",
    volume = "131",
    number = "7",
    pages = "071001",
    year = "2023"
}

@article{Leane:2024bvh,
    author = "Leane, Rebecca K. and Tong, Joshua",
    title = "{Optimal celestial bodies for dark matter detection}",
    eprint = "2405.05312",
    archivePrefix = "arXiv",
    primaryClass = "hep-ph",
    reportNumber = "SLAC-PUB-17773",
    doi = "10.1088/1475-7516/2024/12/031",
    journal = "JCAP",
    volume = "12",
    pages = "031",
    year = "2024"
}

@misc{EarthShineGen,
  author = {{EarthShineGen Developers}},
  title = {EarthShineGen: an event generator for upward-going fermions from dark matter bound in the Earth},
  year = {2026},
  howpublished = {\url{https://github.com/EarthShineGen/EarthShineGen}},
}

@article{Green:2018qwo,
    author = "Green, Adam and Tanedo, Philip",
    title = "{DarkCapPy: Dark Matter Capture and Annihilation}",
    eprint = "1808.03700",
    archivePrefix = "arXiv",
    primaryClass = "hep-ph",
    reportNumber = "UCR-TR-2018-FLIP-L3-37",
    doi = "10.1016/j.cpc.2019.04.010",
    journal = "Comput. Phys. Commun.",
    volume = "242",
    pages = "120--131",
    year = "2019"
}

@article{Feng:2015hja,
    author = "Feng, Jonathan L. and Smolinsky, Jordan and Tanedo, Philip",
    title = "{Dark Photons from the Center of the Earth: Smoking-Gun Signals of Dark Matter}",
    eprint = "1509.07525",
    archivePrefix = "arXiv",
    primaryClass = "hep-ph",
    reportNumber = "UCI-TR-2015-07",
    doi = "10.1103/PhysRevD.93.015014",
    journal = "Phys. Rev. D",
    volume = "93",
    number = "1",
    pages = "015014",
    year = "2016",
    note = "[Erratum: Phys.Rev.D 96, 099901 (2017)]"
}

@article{Gould:1999je,
    author = "Gould, Andrew and Khairul Alam, S. M.",
    title = "{Can heavy WIMPs be captured by the earth?}",
    eprint = "astro-ph/9911288",
    archivePrefix = "arXiv",
    doi = "10.1086/319040",
    journal = "Astrophys. J.",
    volume = "549",
    pages = "72--75",
    year = "2001"
}

@article{IceCube:2016zyt,
    author = "Aartsen, M. G. and others",
    collaboration = "IceCube",
    title = "{The IceCube Neutrino Observatory: Instrumentation and Online Systems}",
    eprint = "1612.05093",
    archivePrefix = "arXiv",
    primaryClass = "astro-ph.IM",
    doi = "10.1088/1748-0221/12/03/P03012",
    journal = "JINST",
    volume = "12",
    number = "03",
    pages = "P03012",
    year = "2017",
    note = "[Erratum: JINST 19, E05001 (2024)]"
}

@article{ParticleDataGroup:2024cfk,
    author = "Navas, S. and others",
    collaboration = "Particle Data Group",
    title = "{Review of particle physics}",
    doi = "10.1103/PhysRevD.110.030001",
    journal = "Phys. Rev. D",
    volume = "110",
    number = "3",
    pages = "030001",
    year = "2024"
}

@article{Achard_2004,
   title={Measurement of the atmospheric muon spectrum from 20 to 3000 GeV},
   volume={598},
   ISSN={0370-2693},
   url={http://dx.doi.org/10.1016/j.physletb.2004.08.003},
   DOI={10.1016/j.physletb.2004.08.003},
   number={1-2},
   journal={Physics Letters B},
   publisher={Elsevier BV},
   author={Achard, P. and others},
   collaboration={L3},
   year={2004},
   month=sep, pages={15--32} }

@article{Highland1975,
  author  = {Highland, Virgil L.},
  title   = {Some Practical Remarks on Multiple Scattering},
  journal = {Nucl. Instrum. Meth.},
  volume  = {129},
  pages   = {497--499},
  year    = {1975},
  doi     = {10.1016/0029-554X(75)90743-0},
  note    = {Erratum: Nucl. Instrum. Meth. {\bf 161} (1979) 171}
}

@article{Lynch:1990sq,
  author  = {Lynch, Gerald R. and Dahl, Orin I.},
  title   = {Approximations to multiple Coulomb scattering},
  journal = {Nucl. Instrum. Meth. B},
  volume  = {58},
  pages   = {6--10},
  year    = {1991},
  doi     = {10.1016/0168-583X(91)95671-Y}
}

@article{Cousins1992,
  author  = {Cousins, Robert D. and Highland, Virgil L.},
  title   = {Incorporating systematic uncertainties into an upper limit},
  journal = {Nucl. Instrum. Meth. A},
  volume  = {320},
  pages   = {331--335},
  year    = {1992},
  doi     = {10.1016/0168-9002(92)90794-5}
}

@article{Read_2002,
  author  = {Read, Alexander L.},
  title   = {Presentation of search results: the $CL_s$ technique},
  journal = {J. Phys. G},
  volume  = {28},
  pages   = {2693--2704},
  year    = {2002},
  doi     = {10.1088/0954-3899/28/10/313}
}

@article{Hamed:2009,
  author       = {Arkani-Hamed, Nima and Finkbeiner, Douglas P. and Slatyer, Tracy R. and Weiner, Neal},
  title        = {A Theory of Dark Matter},
  journal      = {Phys. Rev. D},
  volume       = {79},
  pages        = {015014},
  year         = {2009},
  doi          = {10.1103/PhysRevD.79.015014},
  eprint       = {0810.0713},
  archivePrefix = {arXiv},
  primaryClass = {hep-ph}
}

@article{Pospelov2008,
  author       = {Pospelov, Maxim and Ritz, Adam and Voloshin, Mikhail B.},
  title        = {Secluded WIMP Dark Matter},
  journal      = {Phys. Lett. B},
  volume       = {662},
  pages        = {53--61},
  year         = {2008},
  doi          = {10.1016/j.physletb.2008.02.052},
  eprint       = {0711.4866},
  archivePrefix = {arXiv},
  primaryClass = {hep-ph}
}

@article{Finkbeiner_2009,
  author       = {Cholis, Ilias and Finkbeiner, Douglas P. and Goodenough, Lisa and Weiner, Neal},
  title        = {The PAMELA Positron Excess from Annihilations into a Light Boson},
  journal      = {JCAP},
  volume       = {12},
  pages        = {007},
  year         = {2009},
  doi          = {10.1088/1475-7516/2009/12/007},
  eprint       = {0810.5344},
  archivePrefix = {arXiv},
  primaryClass = {astro-ph}
}

@article{Press:1985ug,
  author  = {Press, William H. and Spergel, David N.},
  title   = {Capture by the Sun of a Galactic Population of Weakly Interacting Massive Particles},
  journal = {Astrophys. J.},
  volume  = {296},
  pages   = {679--684},
  year    = {1985},
  doi     = {10.1086/163485}
}

@article{Silk:1985ax,
  author  = {Silk, Joseph and Olive, Keith A. and Srednicki, Mark},
  title   = {The Photino, the Sun and High-Energy Neutrinos},
  journal = {Phys. Rev. Lett.},
  volume  = {55},
  pages   = {257--259},
  year    = {1985},
  doi     = {10.1103/PhysRevLett.55.257}
}

@article{Krauss:1985aaa,
  author  = {Krauss, Lawrence M. and Srednicki, Mark and Wilczek, Frank},
  title   = {Solar System Constraints and Signatures for Dark Matter Candidates},
  journal = {Phys. Rev. D},
  volume  = {33},
  pages   = {2079--2083},
  year    = {1986},
  doi     = {10.1103/PhysRevD.33.2079}
}

@article{Griest:1986yu,
  author  = {Griest, Kim and Seckel, David},
  title   = {Cosmic Asymmetry, Neutrinos and the Sun},
  journal = {Nucl. Phys. B},
  volume  = {283},
  pages   = {681},
  year    = {1987},
  doi     = {10.1016/0550-3213(87)90293-8},
  note    = {Erratum: Nucl. Phys. B {\bf 296} (1988) 1034}
}

@article{Gaisser:1986ha,
  author  = {Gaisser, Thomas K. and Steigman, Gary and Tilav, Serap},
  title   = {Limits on Cold Dark Matter Candidates from Deep Underground Detectors},
  journal = {Phys. Rev. D},
  volume  = {34},
  pages   = {2206},
  year    = {1986},
  doi     = {10.1103/PhysRevD.34.2206}
}

@article{Gould:1987ir,
  author  = {Gould, Andrew},
  title   = {WIMP Distribution in and Evaporation From the Sun},
  journal = {Astrophys. J.},
  volume  = {321},
  pages   = {560},
  year    = {1987},
  doi     = {10.1086/165652}
}

@article{Gould:1987ju,
  author  = {Gould, Andrew},
  title   = {Resonant Enhancements in WIMP Capture by the Earth},
  journal = {Astrophys. J.},
  volume  = {321},
  pages   = {571},
  year    = {1987},
  doi     = {10.1086/165653}
}

@article{Gould:1987tw,
  author  = {Gould, Andrew},
  title   = {Direct and indirect capture of weakly interacting massive particles by the earth},
  journal = {Astrophys. J.},
  volume  = {328},
  pages   = {919--939},
  year    = {1988},
  doi     = {10.1086/166347}
}

@article{Gould:1991hx,
  author  = {Gould, Andrew},
  title   = {Cosmological density of WIMPs from solar and terrestrial annihilations},
  journal = {Astrophys. J.},
  volume  = {388},
  pages   = {338--344},
  year    = {1992},
  doi     = {10.1086/171156}
}

@article{Kobzarev:1966qya,
  author  = {Kobzarev, I. Y. and Okun, L. B. and Pomeranchuk, I. Y.},
  title   = {On the possibility of experimental observation of mirror particles},
  journal = {Sov. J. Nucl. Phys.},
  volume  = {3},
  pages   = {837--841},
  year    = {1966}
}

@article{Okun:1982xi,
  author  = {Okun, L. B.},
  title   = {Limits of electrodynamics: paraphotons?},
  journal = {Sov. Phys. JETP},
  volume  = {56},
  pages   = {502},
  year    = {1982}
}

@article{Holdom:1985ag,
  author  = {Holdom, Bob},
  title   = {Two U(1)'s and Epsilon Charge Shifts},
  journal = {Phys. Lett. B},
  volume  = {166},
  pages   = {196--198},
  year    = {1986},
  doi     = {10.1016/0370-2693(86)91377-8}
}

@article{Holdom:1986eq,
  author  = {Holdom, Bob},
  title   = {Searching for $\epsilon$ Charges and a New U(1)},
  journal = {Phys. Lett. B},
  volume  = {178},
  pages   = {65--70},
  year    = {1986},
  doi     = {10.1016/0370-2693(86)90470-3}
}

@book{Fabbrichesi:2020wbt,
  author    = {Fabbrichesi, Marco and Gabrielli, Emidio and Lanfranchi, Gaia},
  title     = {The Dark Photon},
  publisher = {Springer},
  series    = {SpringerBriefs in Physics},
  year      = {2021},
  doi       = {10.1007/978-3-030-62519-1},
  eprint    = {2005.01515},
  archivePrefix = {arXiv},
  primaryClass  = {hep-ph}
}

@article{Schuster:2009au,
  author       = {Schuster, Philip and Toro, Natalia and Yavin, Itay},
  title        = {Terrestrial and Solar Limits on Long-Lived Particles in a Dark Sector},
  journal      = {Phys. Rev. D},
  volume       = {81},
  pages        = {016002},
  year         = {2010},
  doi          = {10.1103/PhysRevD.81.016002},
  eprint       = {0910.1602},
  archivePrefix = {arXiv},
  primaryClass = {hep-ph}
}

@article{Schuster:2009fc,
  author       = {Schuster, Philip and Toro, Natalia and Weiner, Neal and Yavin, Itay},
  title        = {High Energy Electron Signals from Dark Matter Annihilation in the Sun},
  journal      = {Phys. Rev. D},
  volume       = {82},
  pages        = {115012},
  year         = {2010},
  doi          = {10.1103/PhysRevD.82.115012},
  eprint       = {0910.1839},
  archivePrefix = {arXiv},
  primaryClass = {hep-ph}
}

@article{Meade:2009mu,
  author       = {Meade, Patrick and Nussinov, Shmuel and Papucci, Michele and Volansky, Tomer},
  title        = {Searches for Long Lived Neutral Particles},
  journal      = {JHEP},
  volume       = {06},
  pages        = {029},
  year         = {2010},
  doi          = {10.1007/JHEP06(2010)029},
  eprint       = {0910.4160},
  archivePrefix = {arXiv},
  primaryClass = {hep-ph}
}

@article{Profumo:2017obk,
  author       = {Profumo, Stefano and Queiroz, Farinaldo S. and Silk, Joseph and Siqueira, Clarissa},
  title        = {Searching for Secluded Dark Matter with {H.E.S.S.}, {Fermi-LAT}, and {Planck}},
  journal      = {JCAP},
  volume       = {03},
  pages        = {010},
  year         = {2018},
  doi          = {10.1088/1475-7516/2018/03/010},
  eprint       = {1711.03133},
  archivePrefix = {arXiv},
  primaryClass = {hep-ph}
}

@article{Bell:2021pyy,
  author       = {Bell, Nicole F. and Dolan, Matthew J. and Robles, Sandra},
  title        = {Solar Gamma Ray Constraints on Dark Matter Annihilation to Secluded Mediators},
  journal      = {Phys. Rev. D},
  volume       = {104},
  pages        = {023024},
  year         = {2021},
  doi          = {10.1103/PhysRevD.104.023024},
  eprint       = {2103.16794},
  archivePrefix = {arXiv},
  primaryClass = {hep-ph}
}

@article{Hiroshima:2025lqm,
  author       = {Hiroshima, Nagisa and Kohri, Kazunori and Paul, Partha Kumar and Sahu, Narendra},
  title        = {Revisiting the limits on dark matter annihilation cross-section and decay lifetime in light of electron and positron fluxes},
  journal      = {JCAP},
  volume       = {08},
  pages        = {048},
  year         = {2026},
  doi          = {10.1088/1475-7516/2026/08/048},
  eprint       = {2510.11700},
  archivePrefix = {arXiv},
  primaryClass = {hep-ph}
}

\end{document}